\documentclass[10pt,twocolumn,floatfix,prl,superscriptaddress,nobibnotes]{revtex4-2}
\usepackage{booktabs}
\newcommand{\mytab}{
\renewcommand{\arraystretch}{1.07}
\begin{tabular}{c|c|c}
\hline
\textbf{Setups} & \textbf{$N_{max}$} &\textbf{\begin{tabular}{@{}c@{}}Fidelity error\\ per photon\end{tabular}
} \\
\hline
Measurement &3 ($\mathcal{F_\text{GHZ}}$=56(2)$\%$)&27(3)$\%$  \\
SIM: $\Delta$=\SI{10}{GHz} & 3 $(\mathcal{F_\text{GHZ}}$=57.1(8)$\%$)&17(2)$\%$\\
SIM: $\Delta$=\SI{30}{GHz} & 4 ($\mathcal{F_\text{GHZ}}$=58.0(8)$\%$)&15(1)$\%$\\
SIM: GaAs QD& 8 ($\mathcal{F_\text{GHZ}}$=51.6(4)$\%$)&8.8(1)$\%$  \\
\hline
\end{tabular}
}

\usepackage{lineno}

\usepackage{graphicx}
\usepackage{dcolumn}

\usepackage{xcolor}

\usepackage{bm}
\usepackage{siunitx}
\usepackage{mathtools}
\DeclarePairedDelimiter\bra{\langle}{\rvert}
\DeclarePairedDelimiter\ket{\lvert}{\rangle}
\usepackage{physics}  
\renewcommand\bra[1]{{\langle{#1}|}}

\usepackage{makecell}

\usepackage{hyperref}
\hypersetup{
    colorlinks=true,
    linkcolor=blue,
    citecolor=red,
    linktoc=all,
}

\begin{document}

\preprint{APS/123-QED}

\title{Deterministic photon source of genuine three-qubit entanglement}
\newcommand{\AffCPH}{Center for Hybrid Quantum Networks (Hy-Q), The Niels Bohr Institute, University~of~Copenhagen,  DK-2100  Copenhagen~{\O}, Denmark}
\newcommand{\AffBasel}{Department of Physics, University of Basel, Klingelbergstra\ss e 82, CH-4056 Basel, Switzerland}
\newcommand{\AffBochum}{Lehrstuhl f\"ur Angewandte Fest\"orperphysik, Ruhr-Universit\"at Bochum, Universit\"atsstra\ss e 150, 44801 Bochum, Germany}
\newcommand{\AffCambridge}{Present address: Cavendish Laboratory, University of Cambridge, JJ Thomson Avenue, Cambridge, CB30HE, United Kingdom}
\newcommand{\AffParis}{Present address: Chimie ParisTech, Université PSL, CNRS, Institut de Recherche de Chimie Paris, 75005 Paris, France}
\newcommand{\AffCorresponding}{Email to: lodahl@nbi.ku.dk}

\affiliation{\AffCPH{}}
\affiliation{\AffBochum{}}

\affiliation{\AffCambridge{}}
\affiliation{\AffParis{}}

\author{Yijian Meng}
\affiliation{\AffCPH{}}
\author{Ming Lai Chan}
\affiliation{\AffCPH{}}
\author{Rasmus B. Nielsen}
\affiliation{\AffCPH{}}
\author{Martin H. Appel}
\affiliation{\AffCPH{}}
\affiliation{\AffCambridge{}}

\author{Zhe Liu}
\affiliation{\AffCPH{}}
\author{Ying Wang}
\affiliation{\AffCPH{}}
\author{Nikolai Bart}
\affiliation{\AffBochum{}}
\author{Andreas D. Wieck}
\affiliation{\AffBochum{}}
\author{Arne Ludwig}
\affiliation{\AffBochum{}}
\author{Leonardo Midolo}
\affiliation{\AffCPH{}}
\author{Alexey Tiranov}
\affiliation{\AffCPH{}}
\affiliation{\AffParis{}}

\author{Anders S. S\o{}rensen}
\affiliation{\AffCPH{}}
\author{Peter Lodahl} 
\email[Corresponding author: ]{lodahl@nbi.ku.dk}
\affiliation{\AffCPH{}}


\date{\today}

\begin{abstract}
Deterministic photon sources allow long-term advancements in quantum optics. A single quantum emitter embedded in a photonic resonator or waveguide may be triggered to emit one photon at a time into a desired optical mode. By coherently controlling a single spin in the emitter, multi-photon entanglement can be realized.  We demonstrate a deterministic source of three-qubit entanglement based on a single electron spin trapped in a quantum dot embedded in a planar nanophotonic waveguide. We implement nuclear spin narrowing to increase the spin dephasing time to $T_2^* \simeq 33$ ns, which enables high-fidelity coherent optical spin rotations, and realize a spin-echo pulse sequence for sequential generation of spin-photon and spin-photon-photon entanglement. The emitted photons are highly indistinguishable, which is a key requirement for scalability and enables subsequent photon fusions to realize larger entangled states. This work presents a scalable deterministic source of multi-photon entanglement with a clear pathway for further improvements, offering promising applications in photonic quantum computing or quantum networks.
\

\end{abstract}

\maketitle



\section{Introduction}

Foundational quantum photonics devices are rapidly developing towards real-world applications for advanced quantum communication and quantum computation~\cite{raussendorf_measurement-based_2003,walther_experimental_2005,zhang_loss-tolerant_2022}. The general scaling-up strategy is to utilize small-size entangled states as seeds that can subsequently be fused by probabilistic linear optics gates to realize a universal resource for quantum-information processing~\cite{rudolph_why_2017,gimeno-segovia_three-photon_2015, bartolucci_fusion-based_2023,uppu2021quantum,lobl2024loss,paesani2023high,istrati2020sequential}. Traditionally these entangled seed states have been generated with probabilistic sources; however, a massive overhead of multiplexing is required to make this approach scalable~\cite{Bartolucci_Passive_2021}.

An alternative and  resource-efficient strategy has been proposed for the deterministic generation of multi-photon entangled seed states using a single quantum emitter~\cite{lindner_proposal_2009,gheri1998entanglement}. Recent experimental work on multi-photon entanglement generation has made significant progress both using single atoms \cite{thomas_efficient_2022} and quantum dots \cite{appel_entangling_2022,cogan_deterministic_2023,coste_high-rate_2023,schwartz2016deterministic}; in the former case an impressive 14-photon high-fidelity deterministic entangled source was realized. Yet solid-state alternatives remain attractive due to their high speed and ease of operation, and potential scalability to multiple emitters \cite{tiranov_collective_2023}. These features are essential when constructing realistic architectures for scaled-up quantum-information processing based on deterministic entanglement sources \cite{uppu2021quantum}.  

Previous work on solid-state emitters has been limited to either two-qubit entanglement \cite{appel_entangling_2022,chen2018highly}, or relied on theoretical assumptions to infer multi-particle entanglement~\cite{schwartz2016deterministic, coste_high-rate_2023, cogan_deterministic_2023}. Ref.\cite{cogan_deterministic_2023}  measured two-qubit correlations and used them to determine the full process map of a single entanglement cycle. Repeated application of the process map was then used to reconstruct an $n$-photon cluster state and infer an entanglement length of ten photons. The validity of this procedure requires that imperfections between different cycles are uncorrelated, i.e., a purely Markovian noise process. Ref.\cite{coste_high-rate_2023}  also measured two-qubit correlations and inferred a lower bound on the three-qubit entanglement fidelity based on two hypotheses: 1) the spin initializes in a perfect mixed state at the beginning of the entanglement sequence, and 2) two successive photons emission having the same polarization after spin initialization. While the assumptions of these studies can be reasonable and are to some extent justified by additional measurements, they lack a direct measurement of the density matrix of the generated three-qubit state. Moreover, a related demonstration was conducted for a source with limited indistinguishability in Ref.~\cite{schwartz2016deterministic}, but it only demonstrated three-partite entanglement by 1.5 standard deviations and did not implement active spin control as required for scalability. 

In this work, we directly verify genuine three-qubit entanglement with a solid-state quantum-dot (QD) source.  High-fidelity realisation of such states has been proposed as the fundamental building block of quantum computing architectures \cite{gimeno-segovia_three-photon_2015}. The generated three-qubit Greenberger–Horne–Zeilinger (GHZ) state consists of one electron spin and two photons. Without any theoretical assumptions or background subtraction, we measure a quantum state fidelity of 56(2)\% and single photon indistinguishability exceeding $90\%$. Further data analysis demonstrates genuine three-qubit entanglement through violation of a biseparablity criteria by 10 standard deviations. 

\section{Results}
We realize a recently proposed time-bin multi-photon entanglement protocol~\cite{lee_source_2019,tiurev_fidelity_2021}. The protocol proceeds by entangling the QD spin with the emission time of deterministically generated single photons~\cite{lee_source_2019,tiurev_fidelity_2021}. Previous works require the natural spin precession at a low magnetic field (few mT)~\cite{schwartz2016deterministic,coste_high-rate_2023,cogan_deterministic_2023}.  For such an approach, scaling up to higher numbers of entangled qubits requires increasing the number of spin precessions within the short spin dephasing time ($\approx$ \SI{2}{\nano\second} for electron spin), but reducing the precession period is at odds with producing indistinguishable photons from two ground states separated by the precession frequency and this constrains the fidelity. In contrast, we operate in the regime of a strong Voigt magnetic field which enables active spin control and clearly distinguishable optical transitions.  The former allows preserving the spin coherence in a noisy solid-state environment via spin-echo sequences and nuclear spin narrowing. The latter is essential for the high photon indistinguishability, enabling fusion operations between different seed states.

\begin{figure}
\includegraphics[scale=1.1]{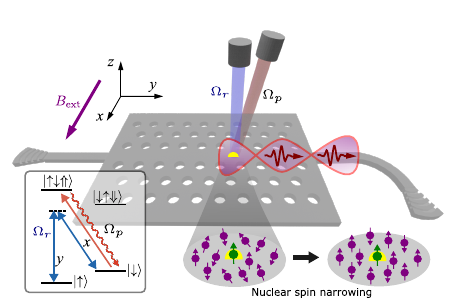}
\caption{\label{fig:fig1_illustration} \textbf{Deterministic GHZ state generation from a QD embedded in a PCW.}  A three-qubit GHZ state is realized by entangling the QD (yellow dot) electron spin and two QD-emitted single photons (red wavepackets). We collect the single photons through the PCW and couple them to free space via a grating outcoupler.  The QD electron spin (dark green arrow) is subjected to the Overhauser field induced by nearby nuclear spins (purple arrows).  To improve $T_2^*$ of the electron spin, we reduce fluctuations of the Overhauser field by employing nuclear spin narrowing. The QD is operated in the Voigt geometry with an in-plane magnetic field of \SI{4}{T} along the $x$-axis. The energy level scheme of the QD is presented in the inset. The Raman laser $\Omega_r$ (blue) is used for coherent spin manipulation of two Zeeman ground states $\ket{\uparrow}$ and $\ket{\downarrow}$. To generate single photons, a picosecond pulsed laser $\Omega_p$ (red) drives photon emissions on the diagonal cycling transition $\ket{\downarrow} \rightarrow \ket{\uparrow\downarrow\Uparrow}$. The same transition is used for the spin readout. Spin initialization is realized through optical pumping on the non-cycling transition $\ket{\downarrow} \rightarrow \ket{\downarrow\uparrow\Downarrow}$. 
}
\end{figure}

In our previous work on deterministic entanglement generation, the entanglement fidelity was primarily limited by the quality of optical spin rotations~\cite{appel_entangling_2022}. There a hole spin was applied since the inhomogeneous spin dephasing time $T^{*}_2 \approx$ \SI{20}{ns} was several times longer than the duration of a $\pi$ pulse on the spin  ($T_p=$~\SI{7}{ns}), which is the key requirement for the protocol. However, the laser rotation pulses were found to induce incoherent $(T_1)$ spin-flip errors of the hole spin, which limited the $\pi$-rotation fidelity  to $F_\pi \approx 88\%$. This error alone sets an upper bound to the two-qubit entanglement fidelity to $77\%$. 

To resolve this bottleneck, we employ the electron spin since the laser-induced spin-flip rate $(\kappa)$ has been shown to be an order of magnitude lower~\cite{bodey_optical_2019}. Conversely, the electron dephasing time is inherently short $T^{*}_2 \approx$ \SI{2}{ns}~\cite{bechtold2015three} and therefore must be prolonged in order to enable high-fidelity spin rotations. To this end,  we implement nuclear spin narrowing by optical coupling~\cite{gangloff_quantum_2019}, see Fig.~\ref{fig:fig1_illustration}, whereby the spin nuclear noise is significantly reduced and $T^{*}_2$ prolonged. The nuclear spin ensemble couples to the electron spin via the Overhauser field, where the shift of the electron spin resonance (ESR) is proportional to the net polarization of the nuclei $I_z$.  In the presence of material strain, optical rotations on the electron spin mediate the nuclear spin-flip transition. By driving and resetting the electron spin with two optical pulses, $I_z$ converges toward a stable point such that its fluctuations are suppressed.

The single-photon emitter is a self-assembled InAs QD embedded in a GaAs photonic-crystal waveguide (PCW) inside a closed cycle cryostat held at 4~K. The QD is grown in a tunable $p$-$i$-$n$ diode heterostructure where a bias voltage charges the QD with an electron as the spin qubit. An external magnetic field of $\textbf{B}_x=4$~T is applied (Voigt geometry), which Zeeman splits the electron spin ground state into $\ket{\uparrow}$ and $\ket{\downarrow}$ by $\Delta_g=22$~GHz and excited state into $\ket{\uparrow\downarrow\Uparrow}$ and $\ket{\downarrow\uparrow\Downarrow}$ by $\Delta_e=12$~GHz, resulting in four optical transitions. 

The PCW offers distinct advantages in the time-bin entanglement protocol. First, the close-to-unity on-chip coupling efficiency ($\beta > 98 \%$ experimentally demonstrated~\cite{arcari2014near}) allows scalable operation towards large photon states, which can be further evanescently coupled to a reconfigurable quantum photonic circuit with a very low loss~\cite{tiecke2015efficient,wang2023deterministic}. Second, the PCW generally Purcell enhances one linear dipole and suppresses the orthogonal dipole, leading to an inherently large optical cyclicity in the $\Lambda$-level scheme, where a spin-preserving optical transition and spin rotating Raman transitions can both be implemented \cite{appel_coherent_2021}.  We find that the decay rate of the inner~(outer) optical transition $\gamma_X$~($\gamma_Y$) to be enhanced (suppressed) and measured a cyclicity of $C=\gamma_X/\gamma_Y=36(3)$, which is the highest recorded value for QDs in the Voigt geometry and an essential figure-of-merit for the entanglement protocol~\cite{tiurev_high-fidelity_2022}.

\begin{figure}[b]
\includegraphics[scale=1.2]{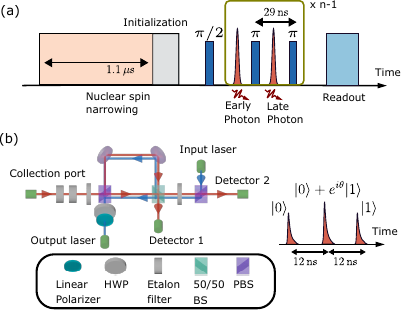}
\caption{\label{fig:schematic_diagram} \textbf{Schematic diagram of time-bin GHZ state generation.} (a) The experimental sequence consists of nuclear spin narrowing and spin initialization, which is then followed by the GHZ protocol with a series of $\pi/2$, $\pi$-pulses (blue) and single photon generations (red). The last part of the sequence (enclosed in the box) is repeated $n-1$ times to generate an $n$ qubit GHZ state. The duration of the full sequence is $1.8~\mu$s. Nuclear spin narrowing consists of two overlapped $1.1~\mu$s Raman pulse and a $1.2~\mu$s pump pulse. The longer pump pulse optically drives the $\ket{\downarrow}\to\ket{\downarrow\uparrow\Downarrow}$ transition and initializes the spin state to $\ket{\downarrow}$ before the GHZ state generation.  (b) Time bin interferometer (TBI) setup: A picosecond pulsed laser (blue) propagates through the long and short arm of TBI to excite a QD emitting a  single photon in the early and late time bin, respectively. The single photon is collected through the same TBI but in the opposite direction. Each photonic qubit has three peaks in the output that are separated by \SI{12}{ns}, which is determined by the path-length difference between the long and short arms of the TBI.}
\end{figure}

The GHZ generation protocol utilizes spin-dependent emission of a single photon, and when combined with coherent spin rotations, it allows entanglement generation between the spin state and the time bin of emitted photons~\cite{appel_entangling_2022}, see Fig.~\ref{fig:schematic_diagram}.  First, a $\pi/2$-rotation pulse prepares the spin in a superposition state $(\ket{\uparrow}-\ket{\downarrow})/\sqrt{2}$. A picosecond laser pulse then drives the cycling transition $\ket{\downarrow}\rightarrow\ket{\uparrow\downarrow\Uparrow}$, see inset of Fig.~\ref{fig:fig1_illustration}, resulting in the single-photon emission in the early time bin. A spin $\pi$-rotation pulse subsequently swaps the $\ket{\uparrow}$ and $\ket{\downarrow}$ population and a second picosecond laser pulse potentially creates a photon in a later time bin. The last $\pi$-pulse swaps spin population again, resulting in the Bell state $(\ket{e\uparrow}-\ket{l\downarrow})/\sqrt{2}$. By iterating this pulse sequence $n-1$ times (apart from the first $\pi/2$-pulse), a GHZ state of $n-1$ photonic qubits and one spin qubit $\ket{\psi_\text{GHZ}}=(\ket{0}^{\otimes n}-\ket{1}^{\otimes n})/\sqrt{2}$ is generated, where we translate $\ket{\uparrow}$($\ket{\downarrow}$) and early(late) emitted photon into the logical qubit $\ket{0}$($\ket{1}$). 

\subsection{Characterization of Spin Control After Nuclear Spin Narrowing}

\begin{figure}[b]  

\includegraphics[trim={1cm 0.2cm 0.2cm 0.3cm},width=0.37\textwidth]{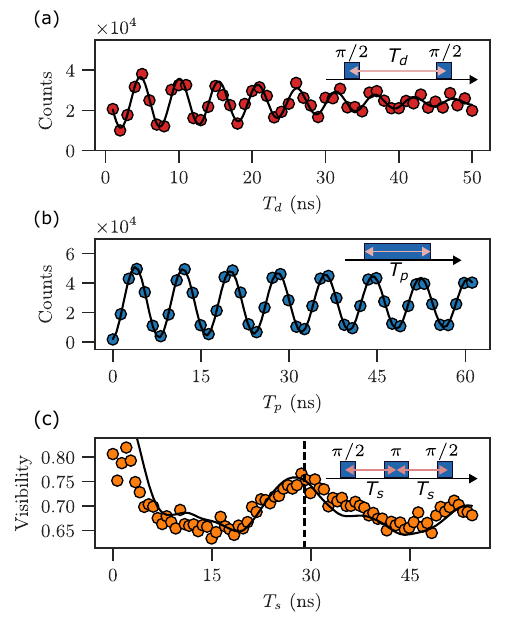}

\caption{\label{fig:Rabi_Ramsey} \textbf{Electron spin spectroscopy.} (a) and (b) display Ramsey interference and Rabi oscillation data, respectively.  We use the former data to extract a spin coherence time of $T_2^*=$~\SI{33(3)}{ns}, where we model the data by assuming inhomogeneous broadening with a Gaussian distribution.   From the Rabi oscillation data, we extract a Q-factor of 34(2).  (c) shows the Hahn-echo visibility as a function of the spacing between the center of $\pi/2$-pulses and the middle $\pi$-pulse. The solid black line is a model of the data by including the nuclear noise spectrum. The dashed-vertical line indicates the echo-pulse spacing used in the GHZ entanglement generation experiment.}
\end{figure}

Prior to running the entanglement protocol, we first characterize the quality of spin control by performing Ramsey spectroscopy and observing Rabi oscillations after nuclear spin narrowing.
In the former, we initialize the electron spin in $\ket{\uparrow}$ by optically pumping $\ket{\downarrow} \rightarrow \ket{\uparrow\downarrow\Uparrow}$ and vary the time delay $T_d$ between two $\pi/2$-rotation pulses. The pulses are \SI{-200}{\mega\hertz} detuned from the spin transitions to reveal oscillations of the Ramsey decay with $T_d$, see Fig.~\ref{fig:Rabi_Ramsey}(a). We model the Ramsey data assuming a Gaussian decay and find $T_2^*=$~\SI{33(3)}{ns},  which is one order of magnitude longer than the spin $\pi$-rotation pulse of $T_p=$~\SI{4}{ns} used in the GHZ generation sequence, thus enabling high rotation fidelity. In Fig.~\ref{fig:Rabi_Ramsey}(b), we drive the spin with a rotation pulse of varying duration $T_p$ to observe Rabi oscillations between $\ket{\uparrow}$ and $\ket{\downarrow}$. We extract a spin Rabi frequency of $\Omega_r=2\pi \cdot 123.6$ \SI{}{MHz} and a quality factor of $\text{Q}=34(2)$, i.e., $\text{Q}$ quantifies the number of $\pi$-pulses before the oscillation visibility falls to $1/e$. 
We estimate the $\pi$-rotation fidelity via $F_{\pi}=\frac{1}{2}(1+e^{-1/\text{Q}})\approx98.6\%$~\cite{bodey_optical_2019}. For comparison, the $\pi$-rotation fidelity error due to finite $T_2^*$ is $2/(\Omega_r T_2^*)^2=0.3\%$~\cite{Appel_thesis_2021}, and we attribute the remaining fidelity error to laser-induced spin-flip processes~\cite{appel_entangling_2022,bodey_optical_2019}: $\frac{1}{2}(1-e^{-\pi\tilde{\kappa}})$. The extracted normalized spin-flip rate is $\tilde{\kappa}=\kappa/\Omega_r\approx3\times 10^{-3}$, which is an order of magnitude lower compared to the case of a hole spin~\cite{appel_entangling_2022}.

To shield the electron spin qubit against nuclear spin noise, the GHZ generation protocol exploits a built-in spin-echo sequence.
We now discuss the spin-echo sequence and benchmarks of the spin coherence. The spin echo consists of three equally spaced $\pi/2$, $\pi$, and $\pi/2$-rotation pulses. We measure the spin echo visibility as a function of the spacing between the pulses. The fringe visibility at each spacing is extracted by altering the phase shift of the last $\pi/2$-pulse between $0$ and $\pi$ relative to the previous pulses. 
We observe a rapid decay of the visibility and a subsequent revival at $T_s\approx$~\SI{29}{ns}, see Fig.~\ref{fig:Rabi_Ramsey}(c). The revival peak is located approximately at the inverse of the Larmor frequency of Indium nuclear spins at \SI{4}{T}~\cite{stockill_quantum_2016}.
The optimal pulse delay corresponding to the peak is then adopted in the entanglement protocol. The maximum spin-echo visibility is 76\%, which constitutes a $20\%$ improvement over previous work~\cite{appel_entangling_2022}. This is an essential figure-of-merit contributing to fidelity of the multi-photon entanglement, as discussed later.

We model the data of Fig.~\ref{fig:Rabi_Ramsey}(c) using the nuclear noise spectral density in InAs QDs as presented by Stockill et al.~\cite{stockill_quantum_2016}. Very good agreement is reached where the observed coherence revival is reproduced.  In the model, we consider the linearly coupled Overhauser field components along the quantization axis of the electron and only use the overall amplitude as a free parameter. We fit the data at $T_s \geq$ \SI{3}{ns}, where the observed discrepancy between data and model at shorter time delays is attributed to interference between the driving pulses. Details of the theoretical model are presented in Supplementary Section VIII.B. The quantitative understanding of the spin-echo visibility is essential for pinpointing how to improve the spin-photon entanglement fidelity in future experiments.

We also perform an extended experiment with two additional $\pi$-pulses at the same spacing, i.e., a pulse sequence of $\pi/2$ - $\pi$ - $\pi$ - $\pi$ - $\pi/2$, and find a maximum echo visibility of 65\%. Compared to the 3-qubit GHZ generation protocol, this measurement is free from excitation errors associated with single-photon generation, the measured visibility therefore poses an upper bound on the three-qubit correlations along the equatorial plane of the Bloch sphere.

\begin{figure*}[t]
\includegraphics[scale=0.47]{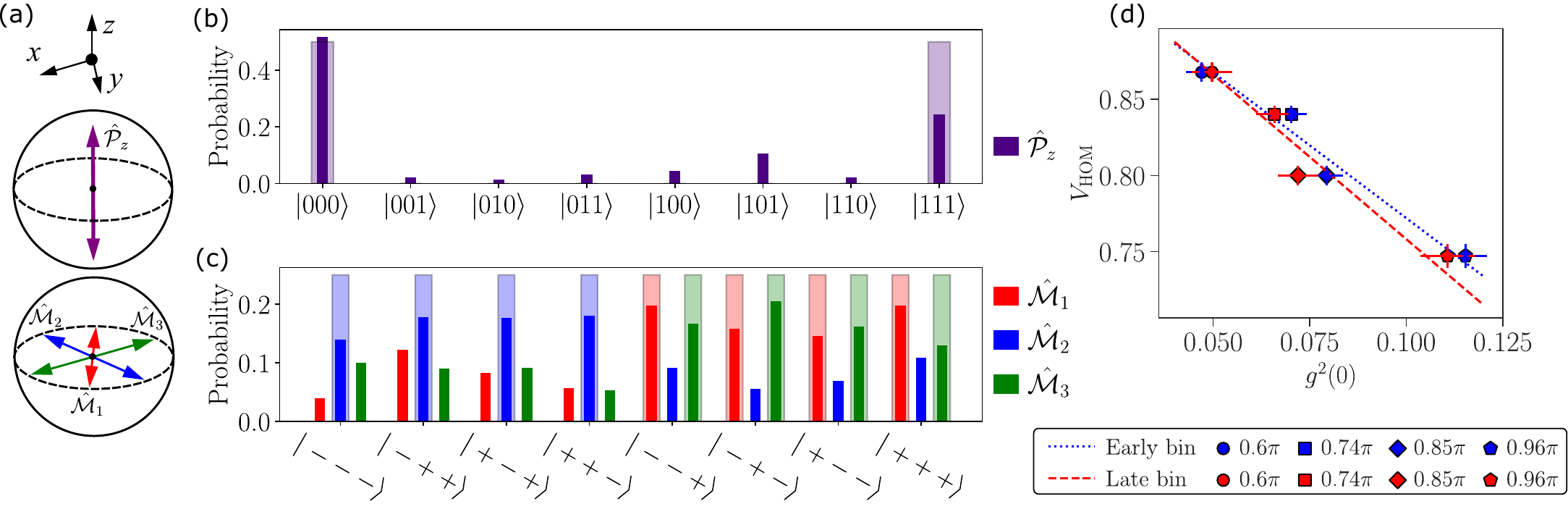}
\caption{\label{fig:bargraph_HOM} \textbf{Fidelity measurement of 3-qubit GHZ state and single-photon characterization.} (a) Illustration of the measurement basis on the Bloch sphere. (b)-(c)  Histogram of fidelity measurement in the $z$-basis and $x,y$-basis, respectively. In (c), $\ket{+}$ and $\ket{-}$ are eigenstates of the operator $\hat{\mathcal{M}}_i$. The acquisition time for each operator is  $\sim$ 2 hours. The shaded bars in (a)-(c) indicate the ideal cases. (d) Measurement of HOM visibility and single-photon purity at different excitation powers (normalized to the $\pi$-pulse power). Single-photon indistinguishability $V_\text{s}$ is extracted from the $y$-intercept of the linear extrapolation. $g^{(2)}(0)$ is measured in the first and third time bins whereas $V_\text{HOM}$ is measured in the second time bin, see main text. The error bars represent one standard deviation.}
\end{figure*}

\subsection{GHZ State Generation and Fidelity Analysis}

The GHZ state is generated and analyzed with the self-stabilized time-bin interferometer in Fig.~\ref{fig:schematic_diagram}(b)~\cite{appel_entangling_2022}. For the photonic qubit, the output consists of three peaks. The first (last) peak corresponds to early (late) emitted photons going through the short (long) arm and constitutes a measurement in the $\ket{1}$ ($\ket{0}$) basis.  The middle peak corresponds to the early photon going through the long arm and the late photon going through the short arm, corresponding to the $\ket{0}+e^{i\theta}\ket{1}$ basis, where $\theta$ is readily controlled via a stack of waveplates and a polarizer~\cite{appel_entangling_2022}. By changing $\theta$, the photonic qubit is measured on the equatorial basis of the Bloch sphere. Spin readout is performed by optically driving the cycling transition $\ket*{\downarrow}\rightarrow\ket*{\uparrow\downarrow\Uparrow}$ with a \SI{200}{ns} pulse.
Since only photons are emitted when the spin state is $\ket{\downarrow}$, an additional $\pi$-pulse is applied before the readout to measure $\ket{\uparrow}$. Similarly, to measure the spin qubit on the equatorial plane, a $\pi/2$ pulse is performed instead.

The entanglement fidelity is decomposed as $\mathcal{F}_{\text{GHZ}}=\Tr\{\rho_{\text{exp}}|\psi_{\text{GHZ}} \rangle \langle  \psi_{\text{GHZ}}|\}=\Tr\{\rho_{\text{exp}}(\mathcal{\hat{P}}_z+\mathcal{\hat{\chi}})/2\}$~\cite{guhne_toolbox_2007}, where $\ket{\psi_{\text{GHZ}}}$ is the ideal GHZ state and $\rho_{\text{exp}}$ is the experimentally measured density matrix. $\mathcal{\hat{P}}_z=\ket{0}\bra{0}^{\otimes n}+\ket{1}\bra{1}^{\otimes n}$ projects all qubits on the $z$-basis and $\mathcal{\hat{\chi}}=\ket{0}\bra{1}^{\otimes n}+\ket{1}\bra{0}^{\otimes n}=\frac{1}{n}\sum_k (-1)^{k} \mathcal{\hat{M}}_k$ is composed of  $\mathcal{\hat{M}}_k$ operators that are equally spread along the $x$-$y$ plane of the Bloch sphere, see Fig.~\ref{fig:bargraph_HOM}(c). The measurement of each operator proceeds from three-photon coincidences between two photonic qubits and a photon from the spin readout in their respective bases.
The normalized three-fold coincidences in each basis are shown in Fig.~\ref{fig:bargraph_HOM}(a)-(b). From here the state fidelity is found to be $\mathcal{F}_{\text{GHZ}}=56(2)$\%, where $\langle\mathcal{\hat{P}}_z\rangle$ and $\langle\mathcal{\hat{M}}_{1,2,3}\rangle$ are 76(2)\% and {$-40(5)$\%, $35(4)$\%, $-33(4)$\%}, respectively.  In the present experiment, the 3-qubit generation rate is estimated to be $R_{\text{GHZ}} = {\eta^2_p} R_{\text{exp}} \simeq$ \SI{70}{kHz}, with the experimental repetition rate of $R_{\text{exp}}=$~\SI{560}{kHz}. The photon efficiency $\eta_p$  includes less-than-$\pi$ optical excitation ($80\%$ due to $0.7\pi$ excitation pulse), emission into the zero-phonon line (95$\%$)~\cite{uppu2020_scalable}, the two-sided waveguide geometry (50$\%$), and waveguide coupling efficiency ($\beta\approx90\%$)~\cite{chan2023chip}. Additional off-chip loss of the entanglement characterization setup constitutes $\approx$\SI{14}{dB} (see Sec. XI of Supplementary Material). Immediate optimization opportunities include operating on a single-sided waveguide~\cite{wang2023deterministic} and increasing the operation speed to $R_{\text{exp}}=$~\SI{1}{MHz} with a reduced nuclear spin narrowing duty cycle, whereby $R_{\text{GHZ}}=$~\SI{0.5}{MHz} is immediately reachable. For subsequent applications of the entanglement source, high-efficiency outcoupling strategies from the chip to the optical fiber can be readily implemented~\cite{wang2023deterministic}.

As a complementary entanglement characterization, we also measure biseparability of the generated state, whereby violation of the following inequality implies genuine three-qubit entanglement:  $|\chi|\leq\sqrt{\rho_{001}\rho_{110}}+\sqrt{\rho_{110}\rho_{001}}+\sqrt{\rho_{101}\rho_{010}}$~\cite{guhne_separability_2010}.  For our experiment, the left and right hand side of the inequality equates to $\langle\mathcal{\hat{\chi}}\rangle=35.8(2.5)\%$ and 9.6(1.1)\%, respectively. The inequality is violated by 10 standard deviations, providing a clear demonstration of genuine 3-particle  entanglement. 
\subsection{Fidelity error contribution and future outlook}

To unravel experimental imperfections, we have performed detailed Monte-Carlo simulations including all relevant physical errors, see Supplementary Sections VIII and IX for full details. For the 3-qubit GHZ state, we predict $\mathcal{F}^{\text{sim}}_{\text{GHZ}}=57.1(8)\%$, which is in very good agreement with the recorded experimental value. Table~\ref{table:1} lists the three main sources of fidelity error. The largest contribution originates from optical excitation of the unwanted cycling transition $\ket{\uparrow}\to\ket{\downarrow\uparrow\Downarrow}$, since the two cycling transitions (the diagonal transitions in Fig.~\ref{fig:fig1_illustration}) are only frequency split by $\Delta=$~\SI{10}{GHz}, which is similar to the excitation laser bandwidth (\SI{12.5}{GHz}). To minimize off-resonant excitation, we excite via an optical $0.7\pi$ pulse at a frequency \SI{2}{GHz} red-detuned from the targeted cycling transition $\ket{\downarrow}\to\ket{\uparrow\downarrow\Uparrow}$. For future experiments, this excitation error could be almost entirely mitigated by orienting the waveguide in the orthogonal direction so that the vertical transitions become cycling. 

The second largest error contribution is the high-frequency nuclear spin noise that was responsible for the reduced spin-echo visibility in Fig.~\ref{fig:Rabi_Ramsey}(c), which then manifests as fidelity error in the entanglement protocol. The noise spectrum is strain dispersion dependent and therefore will constitute a fundamental limit for strain-induced grown QDs (Stranski-Krastanov method) such as InAs QDs in GaAs, but droplet-epitaxy GaAs QDs have proven to be significantly less noisy~\cite{zaporski2023ideal}. Finally, the third main contribution stems from rotation errors due to laser-induced spin flips. A lower rotation error has been observed in a GaAs QD device~\cite{zaporski2023ideal} and the lack of a blocking barrier in these devices suggests a similar design to mitigate laser-induced spin flips in InAs QDs. 

\begin{figure}
\includegraphics[scale=0.55]{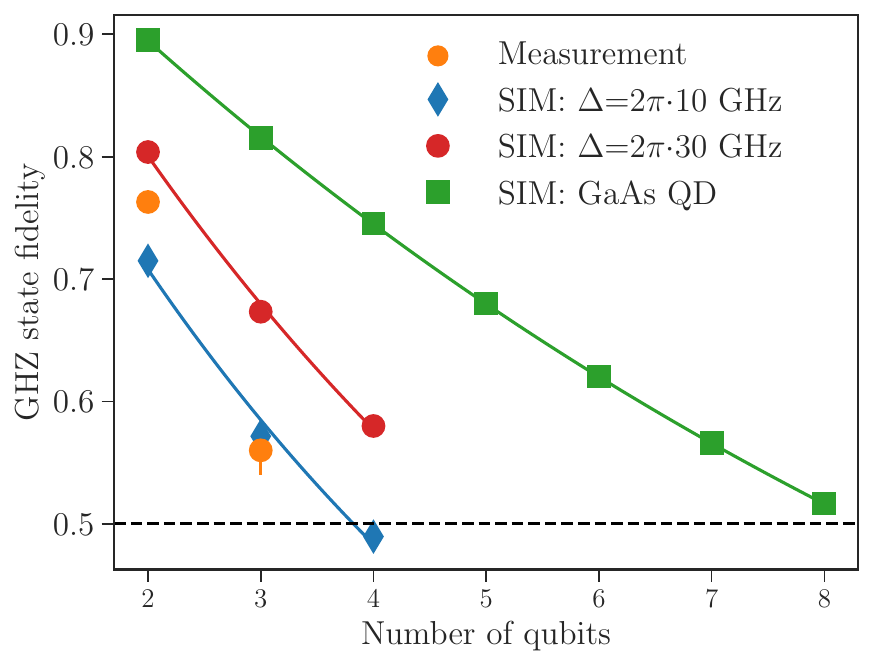}
\\
 \vspace*{0.13cm}

\mytab
\caption{ \textbf{ Summary of experimental results and outlook for future experiments.} Results of three Monte Carlo simulations (SIM): with the current splitting between cycling transitions $\Delta=2\pi~ \cdot $~\SI{10}{GHz}, with the expected modification of a larger $\Delta=2\pi~ \cdot $~\SI{30}{GHz} corresponding to properly oriented waveguides relative to the dipole axes, and with GaAs QD where the effect of nuclear spin noise is negligible and we assume $\Delta=2\pi~\cdot $~\SI{30}{GHz} and a $\pi$ spin rotation fidelity improved by a factor of 2~\cite{zaporski2023ideal}. For all parameters, we run Monte Carlo simulation ($10^4$ realizations) from 2 to 4 qubits, while the fidelity of higher qubit numbers is extrapolated with an exponential fit. For GaAs QD, $N_{max}$ is the predicted maximum attainable number of entangled qubits from the fit, given that $\mathcal{F}_{\text{GHZ}}$ is one standard deviation above 50$\%$. The minor deviation between measurement and simulation is likely due to an overestimation of the nuclear spin noise and an underestimation of other experimental errors, e.g., an imperfect excitation laser polarization.}
\label{fig:outlook}
\end{figure}

In future experiments, we can readily eliminate the first and the most dominant error source by rotating the PCW by $90^\circ$, leading to a Purcell-enhanced $y$-dipole, see Fig.~\ref{fig:fig1_illustration} inset. In this configuration,  the two transitions with the highest energy difference are cycling transitions with a frequency splitting of $\approx$~\SI{30}{GHz}. Consequently, this would allow generating 3 and 4-qubit GHZ states with fidelity of 67\% and 58\%, respectively  (See Sec. VIII of Supplementary Material). Improving beyond this would require reducing the two latter errors. To this end, GaAs QDs have proven to be a highly promising platform. From the numerical simulations, we predict $8.8(1)$\% fidelity error per photon for GaAs QDs, meaning that GHZ strings of 8 qubits should be attainable, where we have used the QD parameters reported in Ref.~\cite{zaporski2023ideal}. Figure \ref{fig:outlook} summarizes the experimental and simulated findings of the GHZ state fidelity for the various experimental situations.

\begin{table}[ht]
\centering
\renewcommand{\arraystretch}{1.07}

\begin{tabular}{c|c}

\hline
\textbf{Primary error sources} & \textbf{Fidelity error} \\
\hline
Off-resonant excitation & 11.4 $\%$ \\
Nuclear spin noise & 6.0 $\%$  \\
Spin-flip error during rotation & 3.2 $\%$  \\
\hline
\end{tabular}
\caption{ Fidelity error contribution of the three primary simulated errors. The value of the fidelity error contribution is calculated as the difference between the fidelity with and without the error source.}
\label{table:1}
\end{table}

\subsection{Single-Photon Purity and Indistinguishability Measurements}
Finally, to characterize the quality of emitted photons, we perform Hanbury-Brown–Twiss (HBT) and Hong–Ou–Mandel (HOM) interferometry to measure single-photon purity and indistinguishability, respectively \cite{appel_coherent_2021}. Here we use a different experimental sequence where we initialize the spin in $\ket{\downarrow}$ with a $\pi$-rotation pulse, and excite the QD twice with the pulsed laser. The pair of emitted photons propagate through TBI and arrive at two detectors in the early, middle and late time bins, which align to the three peaks in Fig.~\ref{fig:schematic_diagram}(b). For HBT, we look at two-detector coincidences in the early or late time bin. We extract $g^{(2)}(0)$ from the ratio between coincidence counts occurring in the same experimental cycle and two consecutive cycles. For HOM, we extract the HOM visibility $V_{\text{HOM}}$ by recording two-detector coincidence counts in the middle bin, normalized to coincidences between the middle bin in one detector and early or late time bin in the other detector. We measure both $V_{\text{HOM}}$ and  $g^{(2)}(0)$ for a range of excitation powers, see Fig.~\ref{fig:bargraph_HOM}(d).

The intrinsic single-photon indistinguishability $V_{\text{s}}$ would be measured in HOM interferometry if the multi-photon contribution and laser leakage were negligible, i.e., $g^{(2)}(0)=0$. We observe a close to linear relationship between $V_{\text{s}}$ and $g^{(2)}(0)$. Using linear regression, i.e., $V_{\text{HOM}}=V_{\text{s}}-F\cdot g^{(2)}(0)$~\cite{ollivier_hong-ou-mandel_2021}, we obtain $F=2.0(3)$ and $V_{\text{s}}=97(2)\%$ after averaging over the data acquired at the first and last time bins.  While $F=2$ has been widely used in the literature~\cite{wang2019towards,somaschi2016near},  the relationship between $g^{(2)}(0)$ and $V_{\text{s}}$ can be more complicated where $F$ varies between 1 and 3, depending on the origin of the multi-photon contribution~\cite{gonzalez_ruiz_single_photon_2023}. $F\approx 2$ indicates that the recorded $V_{\text{HOM}}$ is primarily limited by distinguishable multi-photon component, which could potentially be removed at the cost of lower brightness, e.g., by time gating. To reduce laser leakage, we perform additional frequency filtering using a pair of etalon filters (with a full width at half maximum of \SI{3}{\giga\hertz}) before the TBI. We note the indistinguishability has a lower bound of $92(1)\%$ when $F=1$. The high degree of indistinguishable photons means that the demonstrated entanglement source is readily applicable for realizing photon fusion, which is the essential operation enabling photonic quantum computing~\cite{bartolucci_fusion-based_2023}.

\section{Discussion}

We have presented a direct demonstration of on-demand generation of a genuine three-qubit entangled state using a QD spin-photon interface. This realizes a foundational building block for scalable photonic quantum-information processing towards photonic quantum computers \cite{bartolucci_fusion-based_2023} or one-way quantum repeaters~\cite{borregaard2020one}. The full account of entanglement infidelities is laid out, which is essential for further improvements of the approach in future experiments. 

The InAs QD platform is predicted to be scalable to strings of 4 entangled qubits (Fig.~\ref{fig:outlook}), which suffices for implementing small-scale measurement-based quantum algorithms. Importantly, the high degree of indistinguishability means that such a source can be used to fuse much larger entangled states by de-multiplexing the photons~\cite{hummel2019efficient} and subsequently interfering them in, e.g., an integrated photonic circuit. Interestingly, the current limitation to fidelity can be overcome with a straightforward correction of the device orientation, thanks to a larger splitting between two cycling transitions $\Delta$. We expect the next iteration of the experiment to go to 3 photons and 1 spin, limited by the rotation fidelity error and nuclear spin noise. Longer strings of entangled photons may be realized with strain-free GaAs droplet-epitaxy QDs, where the $\pi$-spin rotation error is halved and the high frequency nuclear noise with a narrower spectrum can be effectively filtered~\cite{zaporski2023ideal}. Such deterministic entanglement sources relying on spin-photon interfaces may offer fundamentally new opportunities for resource-efficient photonic quantum computing architectures \cite{paesani2023high,lobl2024loss}. 

\section*{Acknowledgments}
The authors thank Urs Haeusler, Dorian Gangloff, Mete Atatüre, Giang Nam Nguyen, and Matthias C. Löbl for valuable discussions. We gratefully acknowledge financial support from Danmarks Grundforskningsfond (DNRF 139, Hy-Q Center for Hybrid Quantum Networks), Styrelsen for Forskning og Innovation (FI) (5072-00016B QUANTECH), the European Union’s Horizon 2020 research and innovation program under Grant Agreement No. 820445 (project name Quantum Internet Alliance), the European Union’s Horizon 2020 Research and Innovation programme under Grant Agreement No. 861097 (project name QUDOTTECH), the European Union’s Horizon 2021 Marie Skłodowska-Curie grant No. 101060143 (project name ODeLiCs).

\section*{Author Contributions}

The experiments were conceived by P.L and A.S. and carried out by Y.M., M.L.C, R.N, M.A. The data analysis was performed by Y.M., M.L.C, R.N with support from A.S and P.L. The experimental simulation was carried out by Y.M., M.L.C. and R.N.  The quantum dot device was fabricated by  N.B, A.L, A.W.  The photonic structure was designed and fabricated by A.T, Z.L, Y.W, and L.M.  The paper was written by Y.M, M.L.C, P.L and A.S.

\section*{Ethics declarations}
\subsection{Competing interests}
P.L. is a founder of the company Sparrow Quantum which commercializes single-photon sources. The authors declare no other conflicts of interest.

\section*{Data availability}
Data can be found in the following repository:

https://erda.ku.dk/archives/89bd3313cade9bf037f936f75420d09e/published-archive.html

\bibliographystyle{apsrev4-1} 
\bibliography{Paper3qubits} 

\end{document}


\title{Supplementary Materials - Deterministic photon source of genuine three-qubit entanglement
}

\setcounter{equation}{0}
\setcounter{figure}{0}
\setcounter{table}{0}
\setcounter{page}{1}
\makeatletter
\renewcommand{\figurename}{\textbf{Supplementary Figure}}
\renewcommand{\tablename}{\textbf{Supplementary Table}}
\renewcommand{\refname}{Supplementary References}
\renewcommand{\bibnumfmt}[1]{[S#1]}
\renewcommand{\citenumfont}[1]{S#1}
\renewcommand{\@seccntformat}[1]{%
  \csname the#1\endcsname.\quad
}

\newcommand{\AffCPH}{Center for Hybrid Quantum Networks (Hy-Q), The Niels Bohr Institute, University~of~Copenhagen,  DK-2100  Copenhagen~{\O}, Denmark}
\newcommand{\AffBasel}{Department of Physics, University of Basel, Klingelbergstra\ss e 82, CH-4056 Basel, Switzerland}
\newcommand{\AffBochum}{Lehrstuhl f\"ur Angewandte Fest\"orperphysik, Ruhr-Universit\"at Bochum, Universit\"atsstra\ss e 150, 44801 Bochum, Germany}
\newcommand{\AffCambridge}{Present address: Cavendish Laboratory, University of Cambridge, JJ Thomson Avenue, Cambridge, CB30HE, United Kingdom}
\newcommand{\AffParis}{Present address: Chimie ParisTech, Université PSL, CNRS, Institut de Recherche de Chimie Paris, 75005 Paris, France}
\newcommand{\AffCorresponding}{Email to: lodahl@nbi.ku.dk}

\affiliation{\AffCPH{}}
\affiliation{\AffBochum{}}

\affiliation{\AffCambridge{}}
\affiliation{\AffParis{}}

\author{Yijian Meng}
\affiliation{\AffCPH{}}
\author{Ming Lai Chan}
\affiliation{\AffCPH{}}
\author{Rasmus B. Nielsen}
\affiliation{\AffCPH{}}
\author{Martin H. Appel}
\affiliation{\AffCPH{}}
\affiliation{\AffCambridge{}}

\author{Zhe Liu}
\affiliation{\AffCPH{}}
\author{Ying Wang}
\affiliation{\AffCPH{}}
\author{Nikolai Bart}
\affiliation{\AffBochum{}}
\author{Andreas D. Wieck}
\affiliation{\AffBochum{}}
\author{Arne Ludwig}
\affiliation{\AffBochum{}}
\author{Leonardo Midolo}
\affiliation{\AffCPH{}}
\author{Alexey Tiranov}
\affiliation{\AffCPH{}}
\affiliation{\AffParis{}}

\author{Anders S. S\o{}rensen}
\affiliation{\AffCPH{}}
\author{Peter Lodahl} 
\email[Corresponding author: ]{lodahl@nbi.ku.dk}
\affiliation{\AffCPH{}}

\onecolumngrid

\setcounter{secnumdepth}{3}

\onecolumngrid

\maketitle
{ \onecolumngrid

  \tableofcontents
}

\onecolumngrid
\newpage

\section{Composition of the Quantum Dot Device}
\begin{figure}[htbp]
\includegraphics[scale=0.35]{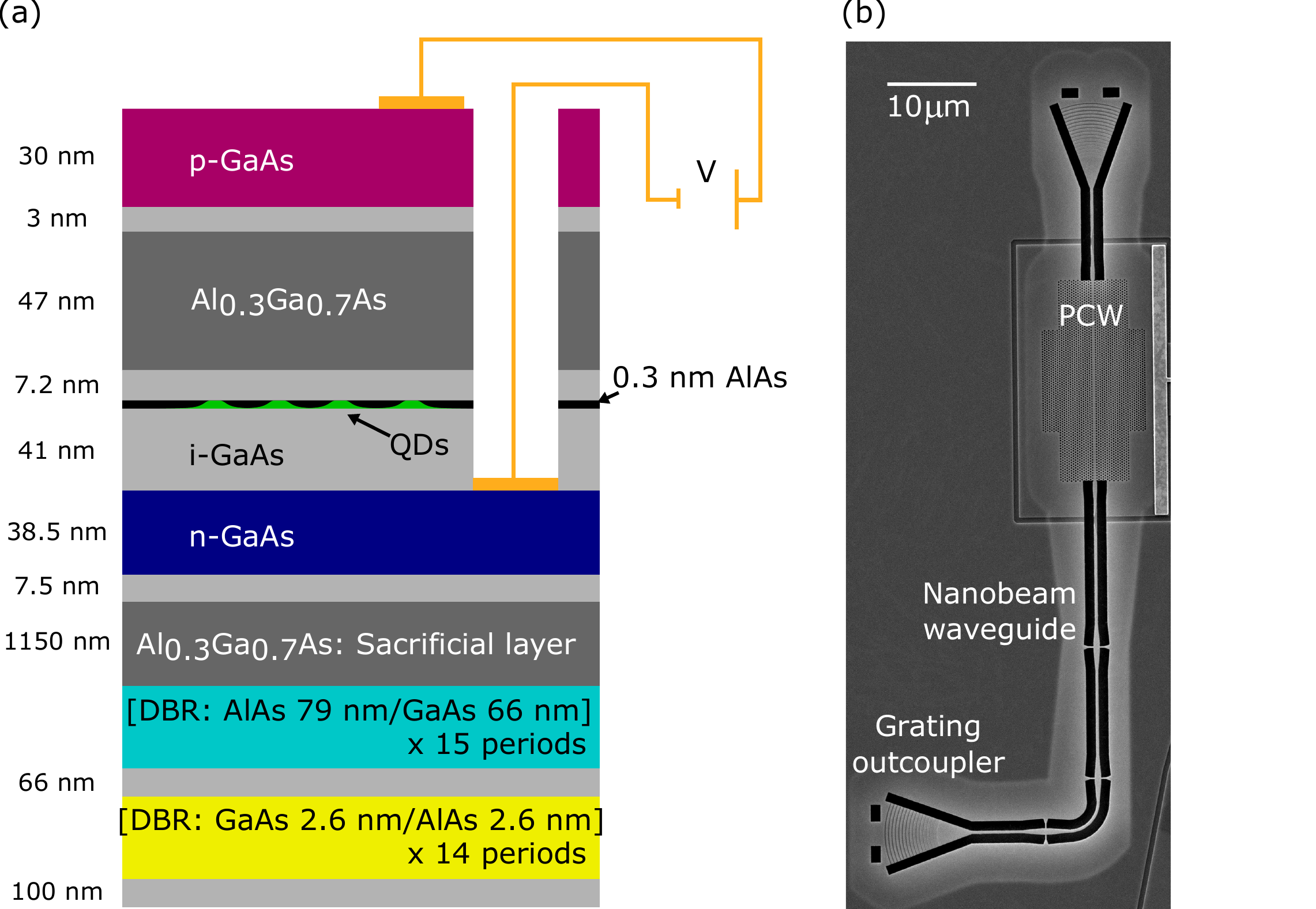}
\caption{(a) Wafer composition of the $p$-$i$-$n$ diode containing self-assembled InAs quantum dots. A $0.3$~nm AlAs layer prevents occupation of the electron wetting layer states~\cite{lobl_wetting_2019}. Applying a bias voltage V across the diode deterministically charges the QD with an electron. (b) Waveguide device used in the experiment. The photonic-crystal waveguide (PCW) uses crystal lattice constant of \SI{246}{nm} and hole radius of \SI{70}{nm} in the slow light region.   }
\label{fig:wafer}
\end{figure}

Our sample is grown by molecular beam epitaxy containing a single layer of self-assembled InAs quantum dots capped by GaAs, embedded in a $p$-$i$-$n$ diode. A forward DC bias voltage V is applied across the diode to stabilize the charge environment, deterministically charge the QD with an electron and tune the frequencies of QD optical transitions via the quantum-confined DC Stark effect~\cite{miller_stark_1984}. Supplementary Fig.~\ref{fig:wafer} outlines the heterostructure of the diode. A $174$ nm-thin GaAs membrane spanning from the hole-rich $p$-doped to the electron-rich $n$-doped layers is grown on top of the $\text{Al}_{0.3}\text{Ga}_{0.7}\text{As}$ sacrificial layer. The layer of InAs QDs (green) is located at the center of the membrane for maximal coupling to the optical field. Detailed fabrication steps of the wafer can be found in Ref.~\cite{uppu2020_scalable}. To fabricate suspended photonic-crystal waveguides (see Supplementary Fig.~\ref{fig:wafer}(b)), the sacrificial layer is first removed using hydrofluoric acid followed by nanostructure etching with electron-beam lithography and reactive ion etching~\cite{midolo_fab_2015,uppu2020_scalable}. Underneath the sacrificial layer is an addition of two stacks of distributed Bragg reflectors (DBR), which increases the reflection of downward-scattered light by $60\%$~\cite{uppu2020_scalable}. This is different than the DBR-absent wafer used in Ref.~\cite{uppu2020_scalable}, and is shown to improve the grating coupler efficiency from $60\%$ to $80\%$~\cite{uppu2020_scalable}. 

\newpage
\section{Optical Imaging Setup and Sample Mount}
\begin{figure}[htbp]
\includegraphics[scale=0.9]{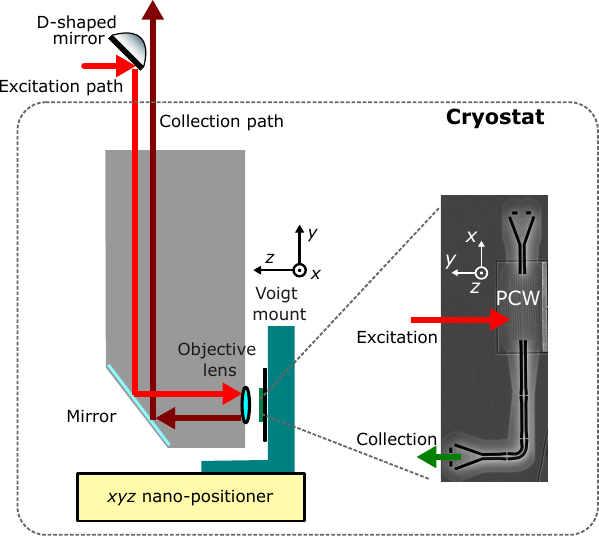}
\caption{Optical imaging setup.}\label{fig:cryo_setup}
\end{figure}
The GaAs photonic chip is placed inside a closed-cycle cryostat (attoDRY1000) at \SI{4}{K}, which contains a vector magnet that generates up to a maximum \SI{5}{T} magnetic field along the $y$-direction and \SI{2}{T} in $x$ (indicated in Supplementary Fig.~\ref{fig:cryo_setup}). To utilize a high magnetic field ($>$~\SI{2}{T}) in the Voigt geometry, the chip is mounted on a custom-made L-shape sample mount (dark green) such that the magnetic field is exerted in-plane relative to the waveguide device. The waveguide structure is brought to
focus by translating 3 piezo nano-positioners beneath the sample mount.

To increase the photon collection efficiency, we use a D-shaped mirror to reflect the excitation light into the microscope objective, which consists of a single lens with a working distance of 1.61 mm and a numerical aperture of 0.68. Due to $\approx70\mu$m spatial separation between QD position and the collection grating coupler in the photonic-crystal waveguide, QD-emitted photons from the coupler are directed towards free space without intercepting on the D-shape mirror and are subsequently collected. Compared to the previous work where a 50:50 beamsplitter was used~\cite{appel_entangling_2022}, the current design improves both the excitation and collection efficiencies by twofold.
\newpage
\section{Electronic Setup for Optical Spin Control}

\begin{figure}[htbp]
\includegraphics[scale=1.1]{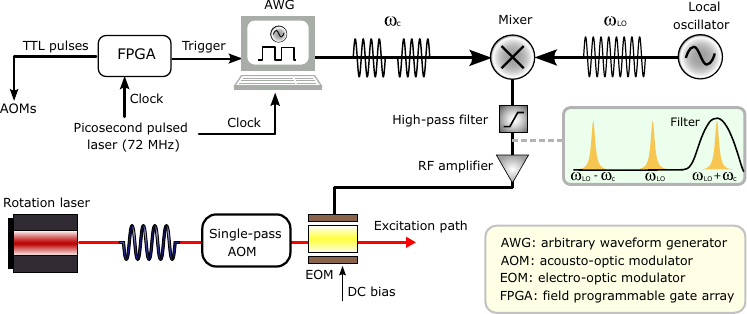}
\caption{Illustration of the electron spin rotation setup. }\label{fig:rotation_setup} 
\end{figure}

Optical spin control pulses are generated by modulating a laser (CTL from Toptica) with a microwave source. The laser propagates through a single-pass acoustic-optic modulator (AOM) setup, and subsequently through an electro-optic modulator (EOM) to generate bi-chromatic nanosecond optical pulses. The EOM is amplitude-modulated by microwave pulses generated using an arbitrary waveform generator (AWG, Active Technologies AWG5064). Since the AWG carrier frequency $\omega_{\text{c}}$ is limited to a maximum \SI{6}{GHz}, it is first mixed with another microwave source (local oscillator in Supplementary Fig.~\ref{fig:rotation_setup}) of frequency $\omega_{\text{LO}}$ using a frequency mixer to match the tens of GHz Zeeman ground-state splitting of the electron spin.

The AWG has an internal IQ mixer which offers flexible control over the relative phase between different pulses. The RF signal out of the AWG is first frequency up-converted via a frequency mixer (Mini-Circuits ZX05-153LH-S+), which creates two frequency sidebands $\omega_{\text{LO}}\pm\omega_{\text{c}}$. Next, The low-frequency band is filtered out using a high band-pass filter (Mini-Circuits ZVBP-10R5G-S+), while the high-frequency band is amplified by a broadband amplifier (ixBlue DR-PL-20-MO). The amplified RF signal is then used to amplitude-modulate the optical signal passing through the EOM (ixBlue NIR-MX800-LN-10), resulting in bi-chromatic pulses (i.e., with two optical sidebands) to drive the two-photon Raman transition. In the current work, $\omega_{\text{c}}=$~\SI{2.8}{GHz} and $\omega_{\text{LO}}=$~\SI{8.2}{GHz} are set to drive an electron Zeeman ground-state splitting
of $\Delta_g =2(\omega_{\text{c}}+\omega_{\text{LO}})=$~\SI{22}{GHz} at \SI{4}{T}. The factor of 2 stems from the frequency difference between two sidebands.


To synchronize between the optical excitation
pulses, spin readout, cooling and rotation pulses in the entanglement pulse sequence, an external RF clock signal of 72.63 MHz generated by a picosecond Ti:Sapphire laser (Coherent
MIRA 900 P) is first divided into two clock signals via a BNC T-splitter to synchronize with the
AWG and a custom-made field
programmable gate array (Cyclone V FPGA from Intel), respectively. The FPGA outputs TTL signals
to produce trigger to the AWG and a time-to-digital
converter (Time Tagger Ultra) as well as square pulses for other AOMs (e.g., for spin readout). The photon detection events are recorded using Time Tagger Ultra and superconducting nanowire single-photon detectors with a timing jitter of 260-$290$~ps.


\newpage
\section{ Entanglement Pulse Sequence and Experimental Procedure}
 In this section, we show explicit steps for generating three-qubit entanglement and elaborate on the experimental procedure used in the experiment. The entanglement sequence for the three-qubit GHZ state can be written in the following:
 
\begin{enumerate}[leftmargin=6em, label={Step \arabic*:}]
\item Initialize the electron spin state in $\ket{\uparrow,\emptyset,\emptyset}$ by optical spin pumping. 
\item  First $\pi/2$ rotation pulse drives the spin to $(\ket{\uparrow,\emptyset,\emptyset}-\ket{\downarrow,\emptyset,\emptyset})/\sqrt{2}$.
\item Optical excitation with a resonant picosecond pulse produces an early photon conditioned on the spin state $\ket{\downarrow}$: $(\ket{\uparrow,\emptyset,\emptyset}-\ket{\downarrow,e,\emptyset})/\sqrt{2}$.
\item A $\pi$ rotation pulse flips the spin ground state to $(-\ket{\downarrow,\emptyset,\emptyset}-\ket{\uparrow,e,\emptyset})/\sqrt{2}$.
\item Another excitation pulse produces a late photon $(-\ket{\downarrow,l,\emptyset}-\ket{\uparrow,e,\emptyset})/\sqrt{2}$.
\item A $\pi$ rotation pulse flips the spin to $(-\ket{\uparrow,l,\emptyset}+\ket{\downarrow,e,\emptyset})/\sqrt{2}$.
\item A resonant excitation pulse produces an early photon $(-\ket{\uparrow,l,\emptyset}+\ket{\downarrow,e,e})/\sqrt{2}$.
\item A $\pi$ rotation pulse flips the spin state to $(\ket{\downarrow,l,\emptyset}+\ket{\uparrow,e,e})/\sqrt{2}$.
\item A resonant excitation pulse produces a late photon $(\ket{\downarrow,l,l}+\ket{\uparrow,e,e})/\sqrt{2}$.
\item A $\pi$ rotation pulse flips the spin state to $(\ket{\uparrow,l,l}-\ket{\downarrow,e,e})/\sqrt{2}$.

\end{enumerate}

Here we use the notation $\ket{\uparrow,\emptyset,\emptyset}$ where the first, second, and third place corresponds to the state of the spin, first photon and second photon, respectively. $\ket{\emptyset}$ denotes the vacuum state. Note that steps 7 to 10 are a repetition of steps 3 to 6. If steps 6 and 10 are each replaced with $\pi/2$ pulse, the final state corresponds to a three-qubit cluster state.

\begin{figure}[htbp]
\includegraphics[scale=0.8]{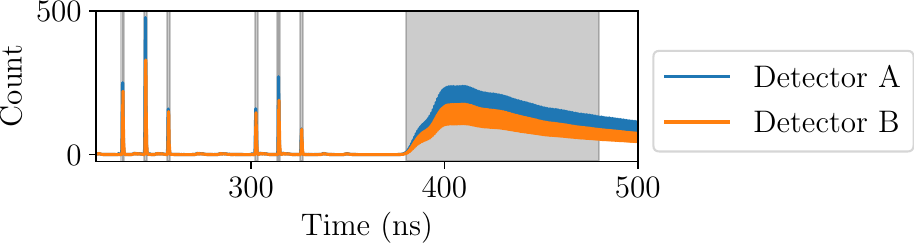}
\caption{\label{fig:Hisgogram} Histogram of the experimental sequence in the first \SI{500}{ns}. The blue and yellow curves correspond to the measurements on the two detectors, respectively. The first~(second) three pulses corresponds to the first~(second) photonic qubit whereas the last long shaded area corresponds to the spin qubit. The rotation pulses (-\SI{650}{GHz}) are filtered out before the detectors using a pair of Etalon filters (\SI{3}{GHz} FWHM).}
\end{figure}

Supplementary Fig.~\ref{fig:Hisgogram} shows the time-resolved histogram of the GHZ measurement sequence. The first (second) three peaks correspond to the detection of the first (second) photonic qubit using the self-stabilized time-bin interferometer. Shaded bars indicate the time gating used for acquiring photon clicks, which we set to be \SI{1}{ns} and \SI{100}{ns} for the photonic and spin qubits, respectively.

To reduce the effect of power drift of the rotation laser at the quantum dot, i.e., pointing instability, we perform interleaving sequences of three-qubit measurement and spin Rabi oscillation. We calculate a scaling ratio between the desired spin Rabi frequency $\Omega^i_r$ for a $4$~ns $\pi$-pulse and Rabi frequency $\Omega^f_r$ fitted from the spin Rabi oscillation. We then adjust the power of the rotation laser in the subsequent iteration by multiplying it by a factor of $\Omega^i_r/\Omega^f_r$. Note that the rotation laser power $P_{\text{rot}}\propto \Omega_r $ since the spin rotation is a two-photon process.

\section{Nuclear Spin Narrowing}

\begin{figure}[htbp]
\includegraphics[scale=0.3]{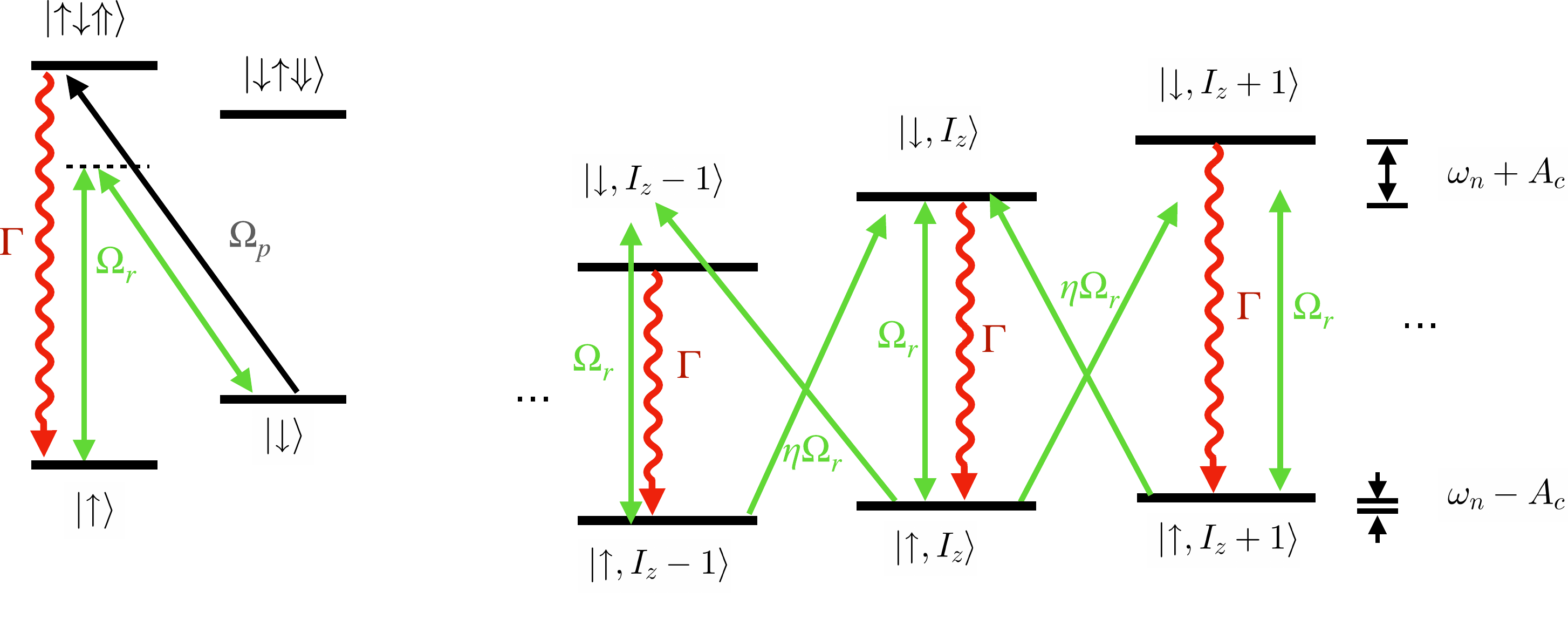}
\caption{\label{fig:nuclear_spin_narrowing} Schematic diagram for nuclear spin cooling. Due to Overhauser shift, the electron spin resonance(ESR) depends on the net nuclear spin polarization $I_z$.} 

\end{figure}

In this section, we explain the mechanism of nuclear spin narrowing process that was also detailed in the literature \cite{gangloff_quantum_2019}. Supplementary Fig.~\ref{fig:nuclear_spin_narrowing} shows the level diagram of a electron spin coupled to the nearby nuclear spin ensemble. The nuclear spin narrowing process is analogous to Raman cooling in cold atom: The Raman laser induces energy exchange between the net nuclear spin polarization $I_z$ and the spin, while another pump laser dissipates the energy of the spin degree of freedom, increasing (decreasing) of $I_z$ when the Raman frequency $\omega_r$ is red(blue) detuned from the electron spin resonance(ESR) $\omega_{\text{ESR}}$. 

The mechanism for nuclear spin narrowing can be further explained in Fig.~\ref{fig:nuclear_spin_narrowing}.  Here the Raman frequency is resonant to ESR at $I_z$ and the spin dissipates from $\ket{\downarrow}$ to $\ket{\uparrow}$ at a rate of $\Gamma$. The ESR is shifted by the Overhauser interaction by 2$A_c$$I_z$, where $A_c$ is the hyperfine coupling energy per nucleus.   When the net nuclear spin polarization reduces to  $I_z-1$,  the net polarization increasing transitions such as $\ket{\uparrow,I_z-1}$ $\rightarrow$  $\ket{\downarrow,I_z}$ become closer to resonant compared to the decreasing transitions such as $\ket{\uparrow,I_z}$ $\rightarrow$ $\ket{\downarrow,I_z-1}$ or $\ket{\uparrow,I_z-1}$ $\rightarrow$ $\ket{\downarrow,I_z-2}$. After further dissipation from $\ket{\downarrow}$ to $\ket{\uparrow}$, the net polarization increases after this process. Similarly, when the net nuclear spin polarization increases above $I_z$, the net polarization decreasing transitions are more resonant and thus in combination with the dissipation channel, the net polarization is reduced. 

During nuclear spin narrowing, the ESR corresponding to $I_z$ becomes a stable point  and the frequency fluctuation around $I_z$ is 
reduced. Therefore, the measured Ramsey time is significantly increased due to the stabilized ESR.


\section{Optical Cyclicity Measurement}

\begin{figure}[htbp]
\includegraphics[scale=0.7]{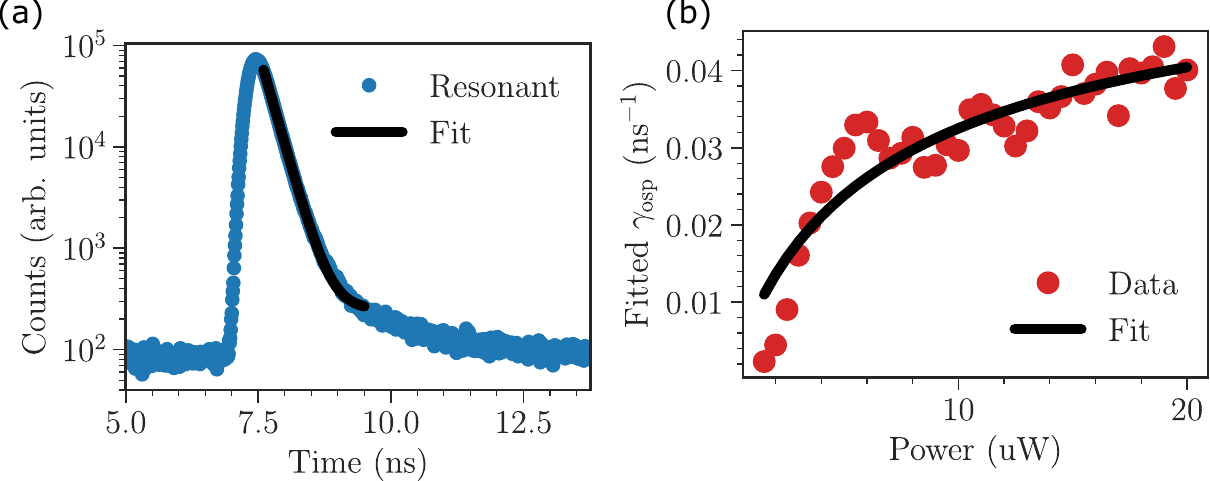}
\caption{\label{fig:CyclicityMeasurement} (a) Time-resolved fluorescence decay of a negatively charged QD measured at \SI{0}{T} magnetic field at bias voltage \SI{1.331}{V}. The QD lifetime is extracted from a single exponential fit. (b) Rate of optical pumping as a function of the pump power.}
\end{figure}
To estimate optical cyclicity of the negatively charged QD in a PCW, we follow the approach in Ref.~\cite{appel_coherent_2021} to perform two sets of measurements shown in Supplementary Fig.~\ref{fig:CyclicityMeasurement}. For the first measurement, we measure its radiative lifetime using $p$-shell excitation by a picosecond Ti:Sapphire laser (Coherent MIRA 900P) locked at $\approx$~\SI{20}{nm} blue-detuned from the QD resonance.  We fit the radiative decay with a single exponential function and extract a lifetime of $1/\Gamma=$~235(1)~\SI{}{ps}, suggesting a waveguide-induced Purcell enhancement of 4.2 compared to $\tau_o\approx$~\SI{1}{ns} in bulk. 

The decay rate $\gamma_Y$ of the inhibited optical transition $\ket{\uparrow\downarrow\Uparrow}\to\ket{\uparrow}$ can then be estimated from fitting its optical pumping rate $\gamma_{\text{osp}}$ as a function of probe power $P$. We perform two-color spin pumping measurements, where a \SI{400}{ns} probe pulse drives the transition $\ket{\downarrow}\to\ket{\uparrow\downarrow\Uparrow}$ with optical Rabi frequency $\Omega_p=\Gamma\sqrt{P/P_{\text{sat}}}$ controlled by $P$ and saturation power $P_{\text{sat}}$. At each power, we fit the fluorescence decay during the probe pulse with a single exponential to extract $\gamma_{\text{osp}}$. From here a list of values of $\gamma_{\text{osp}}$ as a function of $P$ is obtained (Supplementary Fig.~\ref{fig:CyclicityMeasurement}(b)) and subsequently fitted with the function~\cite{appel_coherent_2021}:
\begin{align}
    \gamma_{\text{osp}} (P)=\gamma_Y\int G(\delta_e,\sigma_e) \frac{ P/P_{\text{sat}}}{2P/P_{\text{sat}}+1+4\delta^2_e/\Gamma^2} d\delta_e,
\end{align}
 where $\delta_e$ is the laser detuning from resonance, $G(\delta_e,\sigma_e)$ is a Gaussian distribution of spectral diffusion with a standard deviation in drift of $\sigma_e=2\pi\cdot0.532$~\SI{}{GHz} (extracted independently from the spectral broadening observed in resonant fluorescence spectroscopy). The fit results in $\gamma_Y=0.114(9)~\text{ns}^{-1}$ yielding an optical cyclicity of
 \begin{align}
     C \equiv \frac{\gamma_X}{\gamma_Y} =\frac{\Gamma-\gamma_Y}{\gamma_Y}=36(3),
 \end{align}
 which is $2.5$ times larger than that of the previous QD reported in Ref.~\cite{appel_coherent_2021}.
\section{Experimental Verification of Two-Qubit Entanglement}
As a complementary measurement, we also generate a two-qubit entangled Bell state consisting of an electron spin and a single photon and measure its fidelity. The entangled state is generated by implementing a single repetition of the protocol depicted in Fig.~2a of the main text. Note that the $\pi$-rotation pulse applied after emission of the late time-bin is used only for spin projection in the $-z$-basis. For characterizing spin-photon correlations in the equatorial basis, this rotation pulse is therefore replaced by a $\pi/2$-pulse to form a spin-echo sequence, as has been previously done in Ref.~\cite{appel_entangling_2022}.

The spin-photon entanglement fidelity is exactly decomposed into
$\mathcal{F}_{\text{Bell}}=\Tr\{\rho_{\text{exp}}|\psi_{\text{Bell}} \rangle \langle  \psi_{\text{Bell}}|\}=\Tr\{\rho_{\text{exp}}(\mathcal{\hat{P}}_z+\mathcal{\hat{\chi}})/2\}$~\cite{appel_entangling_2022}, where the ideal spin-photon Bell state is $\ket{\psi_{\text{Bell}}}=\frac{1}{\sqrt{2}}(\ket{e \downarrow}-\ket{l \uparrow})$. $\mathcal{\hat{\chi}}=\frac{1}{2} \sum_k (-1)^{k} \mathcal{\hat{M}}_k=-\mathcal{\hat{M}}_y+\mathcal{\hat{M}}_x$, $\mathcal{\hat{M}}_x=\hat{\sigma}_x$ and $\mathcal{\hat{M}}_y=-\hat{\sigma}_y$. $\mathcal{\hat{P}}_z=\ket{0}\bra{0}^{\otimes 2}+\ket{1}\bra{1}^{\otimes 2}$ measures correlations in the $z$-basis. Conditioned on two-photon coincidences between the photon emission window and spin readout, the probabilities of measuring the spin-photon state in different bases are reported in Supplementary Fig.~\ref{fig:sub_2qubit_bargraph}, with the corresponding expectation values documented in Supplementary Table~\ref{tab:sub_2qubit_table}. The measured fidelity exceeds the classical bound of $50\%$ by $52$ standard deviations, demonstrating two-qubit entanglement.


\begin{figure}[htbp]
\includegraphics[scale=0.8]{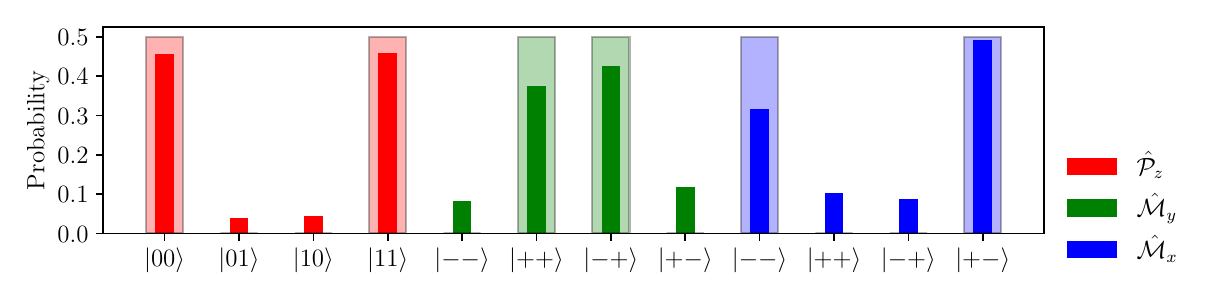}
\caption{\label{fig:sub_2qubit_bargraph} Fidelity characterization of two-qubit entanglement. Shaded bars correspond to the ideal case. }
\end{figure}

\begin{table}[htbp]
    \centering
    \renewcommand{\arraystretch}{1.2}
    \begin{tabular}{c|c}
        \hline
        \textbf{Measurement} & \textbf{Values} \\ 
        \hline
        $\langle\mathcal{\hat{P}}_z\rangle$ & $91.7(4)\%$ \\
        $\langle\mathcal{\hat{M}}_{y}\rangle$ & $-60(1)\%$ \\
        $\langle\mathcal{\hat{M}}_{x}\rangle$ & $62(1)\%$\\
        $\langle\hat{\chi}\rangle$ & $61(1)\%$\\
        \hline
        $\mathcal{F}_{\text{Bell}}$ & $76.3(5)\%$\\
        \hline
    \end{tabular}
    \caption{Summary of expectation values and fidelity for two-qubit entanglement.}
    \label{tab:sub_2qubit_table}
\end{table}

\newpage

\section{Monte Carlo Simulation}
In order to investigate the contributions of various error sources to the state fidelity, we run Monte Carlo simulations taking into account all known errors in the experimental setup. To estimate the future scaling of the entanglement source, we also simulate the fidelity using optimized setups with InAs and GaAs QDs. We verify the validity of the MC simulation by comparing it with analytical expressions derived in Sec. \ref{sec:secMC}.  


\subsection{Simulation Settings and Summary of Results}\label{subsec:MC_details}
Three sets of parameters used for simulations are listed in Supplementary Table~\ref{tab:simFidelityRealParmeter}, corresponding to parameters for the InAs QD in the current experiment, the next iteration with optimized settings (e.g., $90^\circ$-rotated waveguide orientation with $\Delta=$~\SI{30}{GHz}), and future experiments with GaAs QDs.
\begin{table}[!htbp]
    \centering
    \setlength{\tabcolsep}{12pt}
    \renewcommand{\arraystretch}{1.2}
    \begin{tabular}{c|c|c|c}
    \hline
         \multirow{2}{*}{\textbf{Parameters}} & \multicolumn{3}{c}{\textbf{Simulation Setup} } \\
         \cline{2-4}
          & InAs QDs (current) & InAs QDs (optimized) & GaAs QDs \\
         \hline
        Radiative lifetime $1/\Gamma$ & \SI{235}{ps} & \SI{235}{ps} & \SI{235}{ps} \\ 
        Cyclicity $C$ & 36&36&36 \\ 
        Spin rotation Q-factor $\text{Q}$ & 34&36& 68\\
        Spin readout fidelity $F_{\text{r}}$ & 98$\%$& 99$\%$& 99$\%$\\
        Spin initialization fidelity $F_{\text{int}}$ & 99$\%$& 99$\%$& 99$\%$\\
        Pure dephasing rate $\gamma_d$ & \SI{0.069}{}~$\text{ns}^{-1}$& \SI{0.069}{}~$\text{ns}^{-1}$& \SI{0.069}{}~$\text{ns}^{-1}$\\
        Excitation laser detuning & $\SI{-2}{GHz}$& $\SI{0}{GHz}$& $\SI{0}{GHz}$\\
        Frequency splitting between cycling transitions $\Delta$& \SI{10}{GHz}& \SI{30}{GHz}& \SI{30}{GHz}\\
        Excitation pulse shape & Gaussian&Gaussian& Gaussian\\
        Excitation pulse area & 0.7$\pi$& 1$\pi$& 1$\pi$\\
        Excitation pulse duration (FWHM intensity) &  \SI{30}{ps}& \SI{30}{ps}& \SI{30}{ps}\\
        High-frequency nuclear spin noise &   \checkmark & \checkmark & \texttimes\\
                 \hline

    \end{tabular}
    \caption{List of parameters used for the simulation. The excitation laser detuning is the detuning of the excitation laser relative to the targeted cycling transition. The amplitude of nuclear spin noise is extracted from the spin echo visibility measurement, see~Sec.~\ref{subsec:NuclearSpinNoise}. For simulations in GaAs QDs, we assume the nuclear spin noise to be negligible.}
    \label{tab:simFidelityRealParmeter}
\end{table}


In the optimized setup, the increased frequency splitting $\Delta=$~\SI{30}{GHz} due to the rotated waveguide implies that we can use a full $\pi$-pulse for resonant optical excitations, as the probability of off-resonant excitation becomes negligible with a larger $\Delta$. We also expect a modest improvement of the spin readout fidelity to 99$\%$, owing to the larger $\Delta$. For the simulation with GaAs QDs, we take $\Delta=$~\SI{30}{GHz} and assume that the spin rotation Q-factor can be improved by a factor of two with negligible nuclear spin noise. In all cases, we take the pure dephasing rate $\gamma_d\approx0.069~\text{ns}^{-1}$ using the formula $V_s\equiv\Gamma/(\Gamma+2\gamma_d)$~\cite{tighineanu_phonon_2018} where $V_s=96.8\%$ is the single-photon indistinguishability extracted in Fig.~4d of the main text.


To investigate the contribution of experimental imperfections to the state fidelity, we run Monte Carlo simulations using each set of error parameters in Supplementary Table~\ref{tab:simFidelityRealParmeter}. The simulated results for the current experiment are summarized in Supplementary Table~\ref{table:error_contributions}. To isolate the contribution of individual error sources, we first run simulations taking into account all errors, which yields an entanglement fidelity $\mathcal{F}_o$. We then perform a second simulation to extract the fidelity $\mathcal{F}_p$ where an error source $p$ is removed from the simulation. The fidelity error due to $p$ alone can therefore be estimated as $\mathcal{F}_p-\mathcal{F}_0$. Although this approach assumes that all errors are independent, it gives a useful estimate of what to focus on for further optimization.  Apart from three main fidelity error contributions mentioned in Table~1 of the main text, here in Supplementary Table~\ref{table:error_contributions} the corresponding fidelity error from imperfect spin readout, phonon-induced pure dephasing of the excited state and finite optical cyclicity are also included. Note that the simulated fidelity obtained is averaged over $10^4$ Monte Carlo simulations.

\begin{table}[ht]
\centering
\renewcommand{\arraystretch}{1.2}

\begin{tabular}{c|c}

\hline
\textbf{All error sources} & \textbf{Fidelity error} \\
\hline
Off-resonant excitation & 11.4 $\%$ \\
Nuclear spin noise & 6.0 $\%$  \\
Spin-flip error during rotation & 3.2 $\%$  \\
Spin readout error &1.2 $\%$  \\
Phonon-induced pure dephasing & 0.7 $\%$  \\
Finite cyclicity & 0.6 $\%$  \\

\hline
\end{tabular}
\caption{ Simulated fidelity error contribution. The error contribution due to a source $p$ is calculated as the difference in simulated fidelity with and without $p$.}
\label{table:error_contributions}
\end{table}

\newpage

\subsection{Simulation of Nuclear Spin Noise for Spin-echo Visibility Measurement}\label{subsec:NuclearSpinNoise}
\begin{figure*}[h]
    \centering
    \includegraphics[scale=0.45]{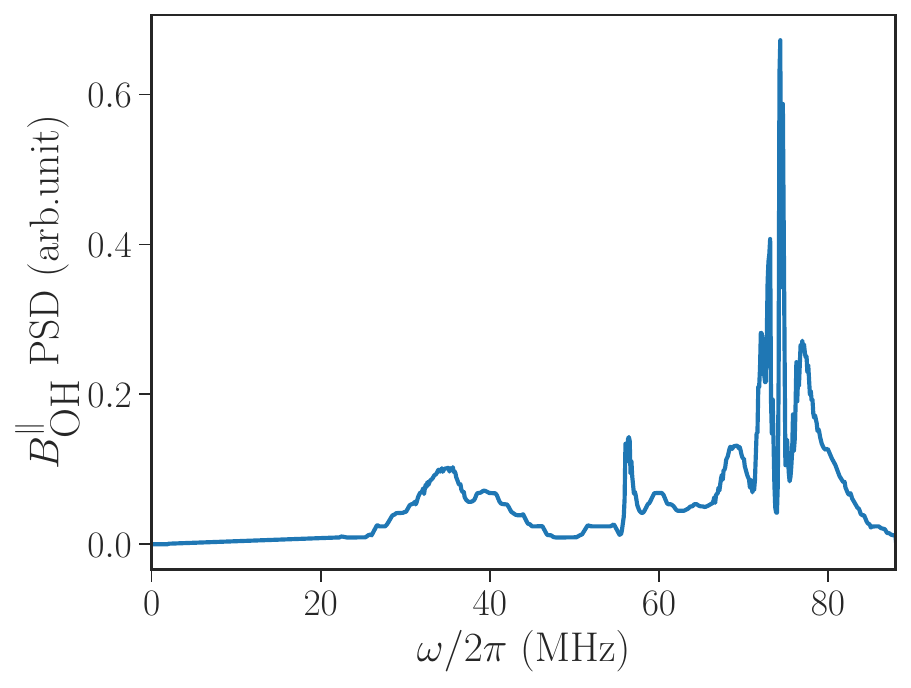}
    \caption{Power spectral density (PSD) of the linearly coupled Overhauser field component along the quantization direction of the electron spin. Here   $\omega$ is the frequency  of the noise arising from the  Larmor precession  of nuclear spins. This plot is reproduced from Stockhill, et al., (2016)~\cite{stockill_quantum_2016}. }
    \label{fig:PSD}
\end{figure*}
To simulate the effect of high-frequency nuclear spin noise in InAs QD on the spin-echo visibility measurement (Fig.~3c of the main text), we first obtain the nuclear noise spectrum by reconstructing the power spectral density (PSD) of the linearly coupled Overhauser field component $B^{\parallel}_{\text{OH}}$ (the superscript ``$\parallel$" denotes field projection along the quantization direction of the electron spin) originating  from In, Ga and As nuclear spins using the data provided in Ref.~\cite{stockill_quantum_2016}. Due to the interplay between Larmor precession of different nuclei species, the linearly coupled Overhauser field noise $\delta B^{\parallel}_{\text{OH}}$ acting on the electron spin also oscillates according to the PSD, resulting in  collapse and revival of the spin echo signal. 
Supplementary Fig.~\ref{fig:PSD} shows the relative amplitude between different nuclear noise frequency components. To simulate the spin echo visibility corresponding to the PSD noise spectrum, we express the phase $\Phi$ induced on the electron spin superposition state $\ket{\uparrow}+e^{i\Phi}\ket{\downarrow}$ in terms of the discretized PSD with a \SI{1}{MHz} resolution:
\begin{align}
    \Phi_j=A\sum_{i} \sqrt{\text{PSD}(\omega_i)}\int^{t_{j+1}}_{t_j}\sin(\omega_i t+\phi_i)dt,
\end{align}
where the square root of the power spectrum $\sqrt{\text{PSD}(w_i)}$ is used to weigh each frequency component $w_i$, and we assign each $w_i$ a random phase $\phi_i$ at the beginning of the simulation. Here the phase $\Phi_j$ is accumulated over the waiting time between $\pi/2$- or $\pi$-rotation pulses, i.e., in a spin echo sequence, where the subscript $j=1$~($j=2$) is the phase accumulated between the first (second) $\pi/2$ pulse at  time $t_1$ ($t_3$) and the middle $\pi$-pulse at time $t_2$. We multiply all frequency components with a common amplitude $A$ which is the only fitting parameter.  
The visibility in the spin echo sequence is measured by the classical contrast in spin readout counts between two opposite phases of the last $\pi/2$-rotation pulse, which corresponds to  $|\cos{(\Phi_2-\Phi_1)}+1|/2$. The zero-frequency component of $\delta B^{\parallel}_{\text{OH}}$ (which determines the spin dephasing time $T^*_2$) is assumed to be filtered out by the spin echo sequence.

To fit the experimental spin echo data in the main text, we fix the visibility at the zero echo delay to 90\%, limited by $98\%$ $\pi$-rotation fidelity and $2\%$ spin initialization error. The obtained fit exhibits strong agreement with the experimental data, specifically demonstrating a rapid decay and subsequent revival of echo visibility as the spin echo spacing increases (see main text Fig.~3). 

\newpage
\section{Comparison of Monte Carlo Simulation with Theory}\label{sec:secMC}

To further validate our simulated results, we now compare the simulated three-qubit fidelity with analytical fidelity expressions for various error sources. For investigating scalability of the time-bin protocol towards more entangled photons, the analytical theory enables generalization of the entanglement fidelity to $N$ photons. The fidelity error per generated photon can also be obtained from the fidelity slope, which is useful for computing the error thresholds of quantum computing architectures. Note that $N=n-1$, where $n$ is the total number of qubit including the QD spin. 

\subsection{Optical Cyclicity}
The GHZ entanglement protocol relies on repeated optical excitations of the same transition to emit single photons of identical frequency and polarization into each time-bin. This requires an optical transition with a strong preferential decay channel which preserves the spin state, and is quantified by the ratio between dominant and inhibited decay rates $C\equiv \gamma_X/\gamma_Y>1$. The effect of finite cyclicity $C$ on the $N$-photon GHZ fidelity has been theoretically investigated in Ref.~\cite{tiurev_fidelity_2021}, where the fidelity was perturbatively expanded to  first order:
\begin{align}
    \mathcal{F}^{(N)}_{\text{GHZ}} \approx 1-\frac{1}{2(C+1)}\bigg(N-\frac{1}{2}\bigg),\label{eq:Cyclicity}
\end{align}
which holds when frequency filtering is applied.
For estimation of the GHZ state fidelity, the  cyclicity error is simulated by introducing a probability of Raman spin-flip process $1/(1+C)$ during optical excitations, where the spin can be flipped by the operator $\ket{\uparrow}\bra{\downarrow}$. 

\begin{figure*}[h]
    \centering
    \includegraphics[scale=0.5]{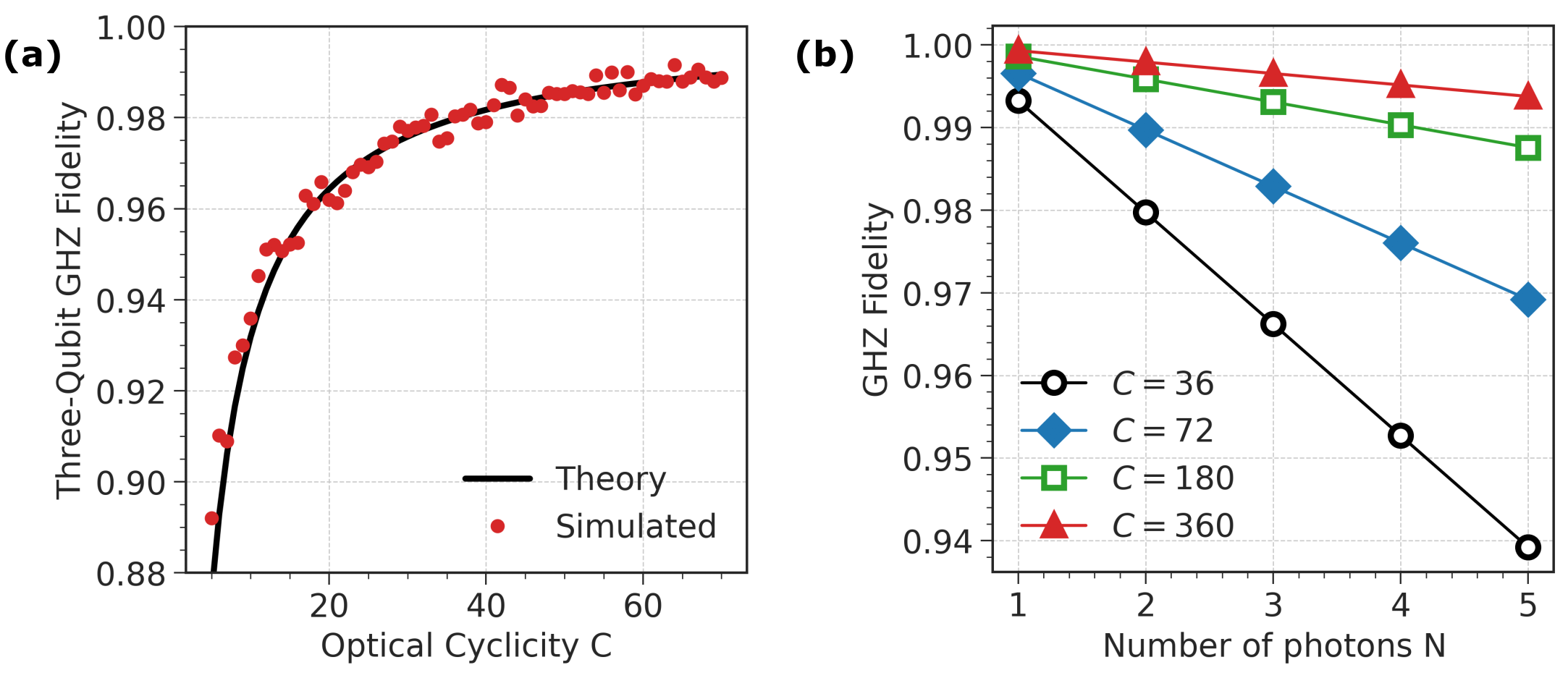}
    \caption{(a) Comparison between the simulated and analytical fidelities under finite cyclicity $C$. (b) Plot of $\mathcal{F}^{(N)}_{\text{GHZ}}$ as a function of photon number $N$ at different values of $C$.}
    \label{fig:compare_c}
\end{figure*}

Supplementary Fig.~\ref{fig:compare_c}(a) plots the simulated three-qubit fidelity $\mathcal{F}^{(2)}_{\text{GHZ}}$ as a function of cyclicity $C$, which agrees well with Eq.~(\ref{eq:Cyclicity}). In Supplementary Fig.~\ref{fig:compare_c}(b) we also plot $\mathcal{F}^{(N)}_{\text{GHZ}}$ versus photon number $N$, at different values of $C$. Black circles correspond to fidelities at the current value of cyclicity $C=36$. A two times improvement in $C$ gives $0.68\%$ fidelity error per photon.

\newpage
\subsection{Incoherent Spin-flip Error}
Spin decoherence induced by the spin rotation laser results in depolarization of the entangled spin-photon state, which has previously been observed in Refs.~\cite{appel_entangling_2022,chan2023chip}. Although the decoherence mechanism is not well-understood, here we model the laser-induced spin decoherence as a $T_1$ process, i.e., the QD electron spin is removed by the red-detuned rotation laser~\cite{lochner_internal_2021} and the QD is later repopulated with a random electron spin state. In our model we assume this process occurs instantaneously. In reality, no photon is detected if the QD spin is absent.  It thus has less influence on the fidelity measurement if the process is not instantaneous.  This could lead to overestimation of the spin-flip error.  
After the QD is repopulated, the resulting spin state is assumed to have an equal probability to end up in either $\ket{\uparrow}$ or $\ket{\downarrow}$, which is a reasonable approximation given $\hbar\Delta_g \ll k_\text{B}T$, where $\Delta_g$ is ground-state splitting, $T$ is the sample temperature, and $\hbar$ and $k_B$ are the reduced Planck constant and Boltzmann constant, respectively.
\begin{figure*}[h]
    \centering
    \includegraphics[scale=1.2]{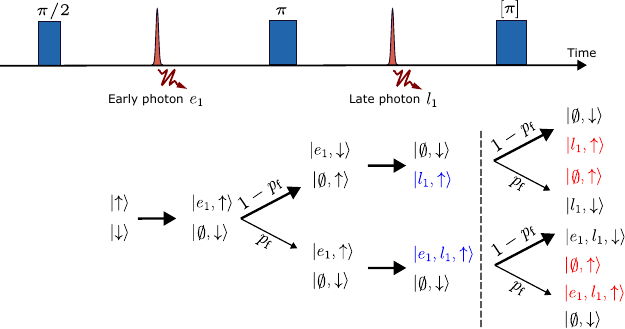}
    \caption{Estimation of $\langle\mathcal{\hat{P}}_z\rangle$ for 2-qubit state using a tree diagram. After each $\pi$-pulse, there are two possible outcomes with probability of spin-flip ($\textit{p}_f$, effective $2\pi$ rotation due to the spin flip and $\pi$-pulse) and no spin-flip ($1-\textit{p}_f$, effective $\pi$-rotation), respectively. The red (blue) colored states are detected using measurement sequences with~(without) the last $\pi$ pulse $[\pi]$. }
    \label{fig:tree_graph_2qubit}
\end{figure*}


To estimate how spin-flip error affects the entanglement fidelity, we analyze all possible spin-flip events after an entanglement sequence and track down the probabilities of individual states after each operation. During each spin $\pi$-rotation, the spin state has a probability of $\textit{p}_f=\frac{1}{2}(1-e^{-\kappa T_\pi})$ to flip its state (on top of the $\pi$ rotation).

As an example, we break down all possible outcomes from the two-qubit entanglement sequence which is shown in a tree diagram  (see Supplementary Fig.~\ref{fig:tree_graph_2qubit}). After the first $\pi/2$-rotation pulse, the resulting state has equal population in $\ket{\uparrow}$ and  $\ket{\downarrow}$, regardless of whether a spin flip occurs. Since $\langle\mathcal{\hat{P}}_z\rangle$  evaluates the population contrast between basis states (no coherence terms required), a spin flip during the first $\pi/2$-pulse does not affect the $\langle\mathcal{\hat{P}}_z\rangle$ measurement. However, during the first $\pi$ rotation: If a spin flip occurs, either no photon is emitted ($\ket{ \emptyset,\downarrow}$) or both an early photon and a late photon are emitted ($\ket{e_1, l_1,\uparrow  }$). The former state does not count toward the overall statistics since no photon is detected, while the latter has twice the probability since photons from both  $\ket{ e_1,\uparrow}$ and $ \ket{ l_1,\uparrow}$ can be detected (our collection efficiency is limited, making it unlikely that we detect both). Furthermore, since only $\ket{\uparrow}$ can be measured by optical spin pumping, an additional $\pi$ pulse is used to transfer $\ket{\downarrow}$ to $\ket{\uparrow}$, which leads to 4 additional possible outcomes.  Summing over the probability of all detected states, the Z-basis expectation value is:
\begin{align}
    \langle\mathcal{\hat{P}}^{(2)}_z\rangle 
    = \frac{\textit{p}_f+(1-\textit{p}_f)+\textit{p}_f^2+(1-\textit{p}_f)^2}{2\textit{p}_f+(1-\textit{p}_f)+2\textit{p}_f^2+(1-\textit{p}_f)^2}\approx 1-\textit{p}_f,
\end{align}
where the first-order expansion in the last step further approximates the results to one spin-flip event per entanglement sequence.

Similarly, the $n$-qubit Z-basis expectation value can be expanded in the first order as
\begin{align}
    \langle\mathcal{\hat{P}}^{(n)}_z\rangle\approx 1-(2n-3)\textit{p}_f,
\end{align}

For $\langle\hat{\chi}\rangle$, when a spin-flip event occurs, the spin coherence drops to zero as the entangled state becomes mixed. Therefore,
\begin{align}
    \langle\hat{\chi}\rangle
    =1\times(e^{-\kappa T_s}) + 0\times(1-e^{-\kappa T_s}) = e^{-\kappa T_s},
\end{align}
where $T_s$ is the total duration of the rotation pulse in a sequence, i.e., $T_s=(2n-3)T_{\pi}$.

To validate our Monte Carlo simulation, we compare the above model with the simulated fidelity in Supplementary Fig.~\ref{fig:compare_kappa_pheno}, which shows good agreement in the limit of $\kappa/\Omega_r\ll 1$. The spin-flip error $\kappa/\Omega_r$ introduces damping to the visibility of Rabi oscillations and is related to the quality factor $\text{Q}$ via the $\pi$-pulse duration $T_\pi=\pi/\Omega_r$:
\begin{align}
    \frac{1}{\text{Q}} \equiv \frac{T_\pi}{T_{1/e}} = \pi\frac{\kappa}{\Omega_r},\label{eq:Qfactor}
\end{align}
which holds when the rotation fidelity error due to finite $T^*_2$ is negligible, i.e., $\Omega_r\gg 1/T^*_2$. As a subject of interest, in Supplementary Fig.~\ref{fig:compare_kappa_pheno}(b), $\mathcal{F}^{(N)}_{\text{GHZ}}$ is plotted in the limit of small photon number $N \le5$ at various values of $\text{Q}$. Here we observe that a twofold reduction in $\kappa$ (or improvement in $\text{Q}$) could minimize the fidelity error per photon to $2\%$.

\begin{figure*}[h]
    \centering

    \includegraphics[scale=0.53]{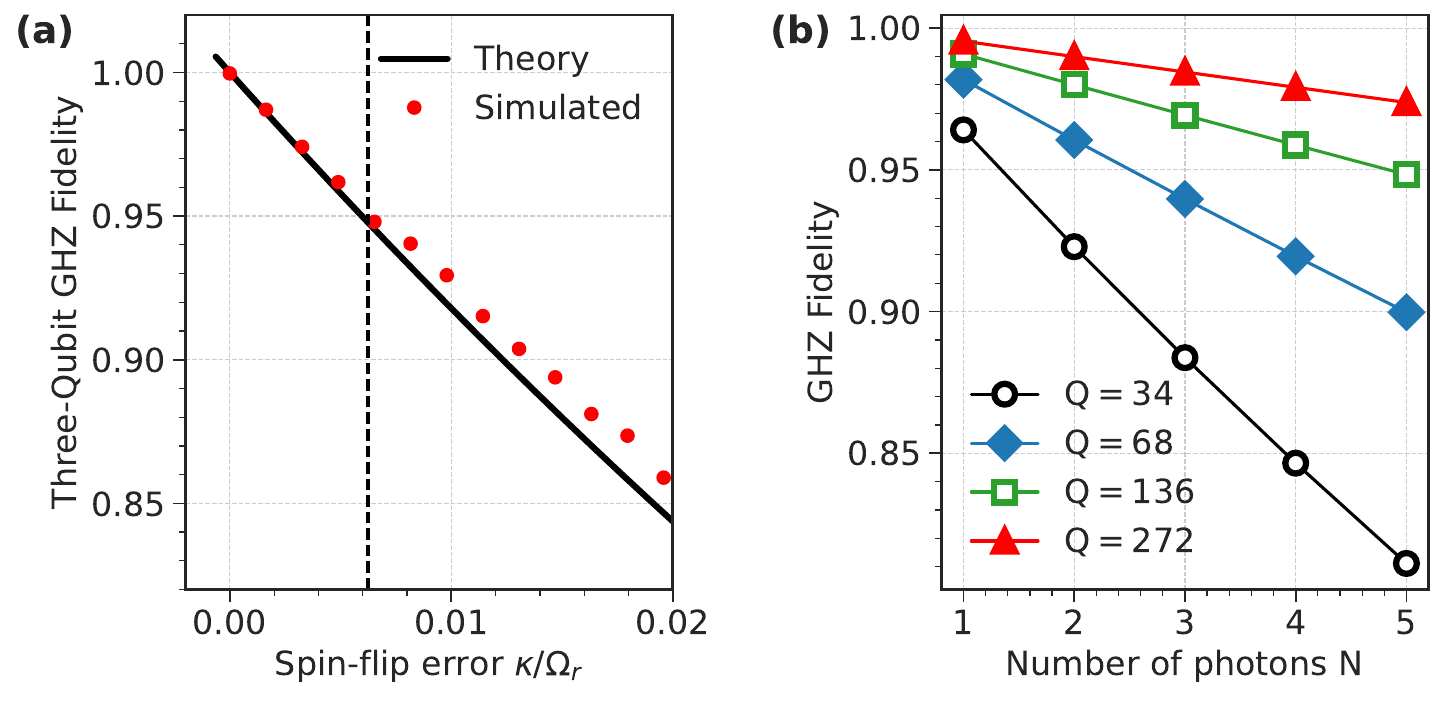}
    \caption{(a) Comparison between the simulated and analytical fidelities as a function of normalized spin-flip rate $\kappa/\Omega_r$. The vertical dashed line corresponding $\text{Q}=34$. (b) Plot of $\mathcal{F}^{(N)}_{\text{GHZ}}$ as a function of photon number $N$ for different quality factor $\text{Q}$.}
    \label{fig:compare_kappa_pheno}
\end{figure*}

\newpage
\subsection{Initialization Fidelity}
\begin{figure*}[h]
    \centering
    \includegraphics[scale=0.5]{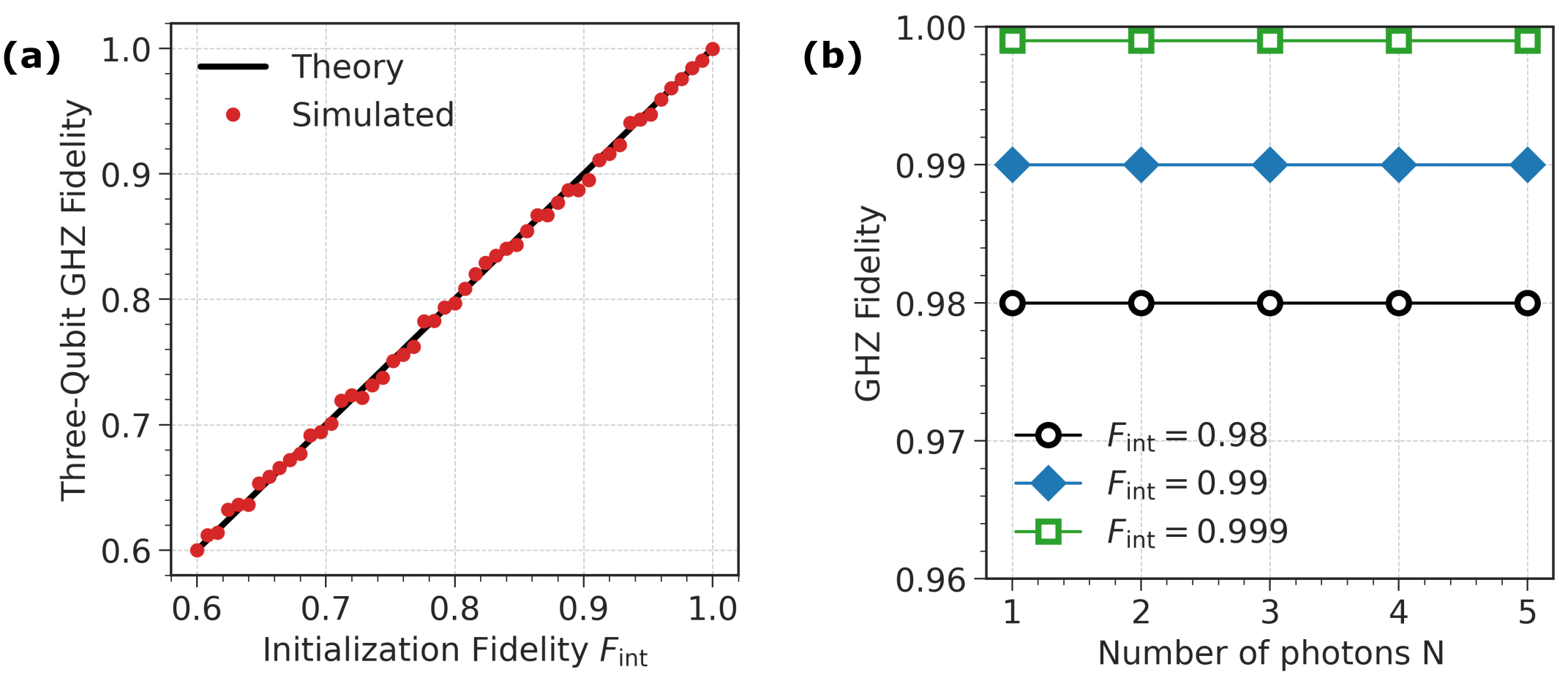}
    \caption{(a) Comparison between the simulated and analytical fidelities under initialization error. (b) Plot of $\mathcal{F}^{(N)}_{\text{GHZ}}$ as a function of photon number $N$ at different values of $F_{\text{int}}$.}
    \label{fig:compare_Finit}
\end{figure*}

To account for initialization error in the simulation, we define $F_{\text{int}}$ as the probability of successful initialization of the spin state $\ket{\uparrow}$ by optical pumping of the transition $\ket{\downarrow}\to\ket{\Downarrow\downarrow\uparrow}$. A random number $i\in [0,1]$ is drawn and compared with $F_{\text{int}}$. If $i\leq F_{\text{int}}$ ($i> F_{\text{int}}$), the initial state becomes $\ket{\uparrow}$ ($\ket{\downarrow}$). Similarly, this error is included in the theory by modifying the spin density matrix $\Tilde{\rho}_s$ as
\begin{align}
    \Tilde{\rho}_s \to F_{\text{int}}\Tilde{\rho}_s + (1-F_{\text{int}}) \rho^{\text{ideal}}_s,
\end{align}
where $\rho^{\text{ideal}}_s$ is the spin density matrix after applying a perfect $\pi/2$ pulse on the spin state $\ket{\downarrow}$. 
It is straightforward to show that the entanglement fidelity is expressed by
\begin{align}
    \mathcal{F}^{(N)}_{\text{GHZ}}=F_\text{int},\label{eq:int}
\end{align}
which does not depend on the photon number in the generated state. Supplementary Fig.~\ref{fig:compare_Finit}(a) compares the simulation and theory, and Supplementary Fig.~\ref{fig:compare_Finit}(b) shows that the initialization error is directly mapped to $\mathcal{F}^{(N)}_{\text{GHZ}}$ as predicted by Eq.~(\ref{eq:int}). 
\newpage
\subsection{Spin Readout Fidelity}

Spin readout is performed in the experiment by optically pumping the $\ket{\downarrow}\to\ket{\Uparrow\downarrow\uparrow}$ transition, where $\ket{\downarrow}$ ($\ket{\uparrow}$) is translated to $\ket{1}$ ($\ket{0}$). To incorporate spin readout error into the simulation, we introduce $F_\text{r}$ as the fidelity of reading out $\ket{1}$, whereas $1-F_\text{r}$ corresponds to the probability of erroneously reading $\ket{0}$.
The entanglement fidelity is:
\begin{align}
    \mathcal{F}^{(N)}_{\text{GHZ}} &= \frac{1}{2}(\langle\mathcal{\hat{P}}_z\rangle+\langle\hat{\chi}\rangle),
\end{align}
where 
\begin{align}
\langle\mathcal{\hat{P}}_z\rangle=\ket{0}\bra{0}^{\otimes{n}}+\ket{1}\bra{1}^{\otimes{n}},\quad\quad\langle\hat{\chi}\rangle=\ket{0}\bra{1}^{\otimes{n}}+\ket{1}\bra{0}^{\otimes{n}}=\frac{1}{n}\sum^{n}_{k=1} (-1)^{k} \langle\mathcal{\hat{M}}_k\rangle.
\end{align}
For $\langle\mathcal{\hat{P}}_z\rangle$, reading out the wrong spin results in the detection of the wrong state, i.e, $\ket{1}^{\otimes N}\ket{0}_s\bra{1}^{\otimes N}\bra{0}_s$ or $\ket{0}^{\otimes N}\ket{1}_s\bra{0}^{\otimes N}\bra{1}_s$. Therefore, $\langle\mathcal{\hat{P}}_z\rangle=F_\text{r}+0(1-F_\text{r})=F_\text{r}$.

On the other hand, the measurement of $\langle\hat{\chi}\rangle$ requires projection of each qubit onto $\mathcal{\hat{M}}_k$ basis with eigenstates $\ket{\pm_k}=(\ket{0}\pm e^{ik\pi/n}\ket{1})/\sqrt{2}$, where $k\leq n$ is the number of qubits. Since in the experiement spin readout is only available in $\ket{1}$, we use two separate measurement configurations for every $\mathcal{\hat{M}}_k$ basis: In one configuration, a spin rotation pulse $\hat{R}(\pi/2,\theta=ik\pi/n)$ is applied to transfer $\ket{+_k}$ ($\ket{-_k}$) to $\ket{1}$ ($\ket{0}$). In the other configuration, $\hat{R}(\pi/2,\theta=ik\pi/n+\pi)$ transfers $\ket{+_k}$ ($\ket{-_k}$) to $\ket{0}$ ($\ket{1}$). Thus, readout of the wrong spin state is equivalent to switching between the $\ket{+}$ and $\ket{-}$ spin states. To see this, $\langle\hat{\chi}\rangle$ can be written as:
\begin{align}
    \langle\hat{\chi}\rangle&=\ket{0}\bra{1}^{\otimes{n}}+\ket{1}\bra{0}^{\otimes{n}}=\frac{(-1)^k}{2^n}\bigg[\bigg((\ket{+_k}-\ket{-_k})({\bra{+_k}+\bra{-_k}})\bigg)^{\otimes{n}}+\bigg((\ket{+_k}+\ket{-_k})(\bra{+_k}-\bra{-_k})\bigg)^{\otimes{n}}\bigg].\label{eq:chi}
\end{align}
After factoring out Eq.~(\ref{eq:chi}), only the diagonal terms have non-zero coefficients of either $1/2^n$ or $-1/2^n$, depending on the parity of the sum of $k$ and the number of $\ket{-_k}$ terms in the state, i.e., for the two-qubit case, $\ket{+_k+_k}\bra{+_k+_k}$ and  $\ket{+_k-_k}\bra{+_k-_k}$ have opposite signs due to their even and odd numbers of $\ket{-_k}$ terms, respectively. Therefore, $\langle\chi\rangle$ is given by the difference between even parity and odd parity terms. By measuring the wrong spin state, i.e., switching between $\ket{+}$ and $\ket{-}$ for the spin qubit, the parity of all diagonal terms is switched with probability $F_{\text{r}}$, leading to $\langle\hat{\chi}\rangle=\mathcal{F_\text{r}}+(-1)(1-\mathcal{F_\text{r}})$.
The overall readout fidelity is then: 
\begin{align}\label{eq:Fread}
\mathcal{F}^{(N)}_{\text{GHZ}}&=\frac{3\mathcal{F_\text{r}}-1}{2},
\end{align}
which is indeed independent of the photon number $N$ in the GHZ state, as spin readout error is always performed at the end of the experimental protocol. In the simulation, the output density matrix is simulated for each measurement setting, and the readout error is introduced as the final step of the simulation. 
\begin{figure*}[h]
    \centering
    \includegraphics[scale=0.5]{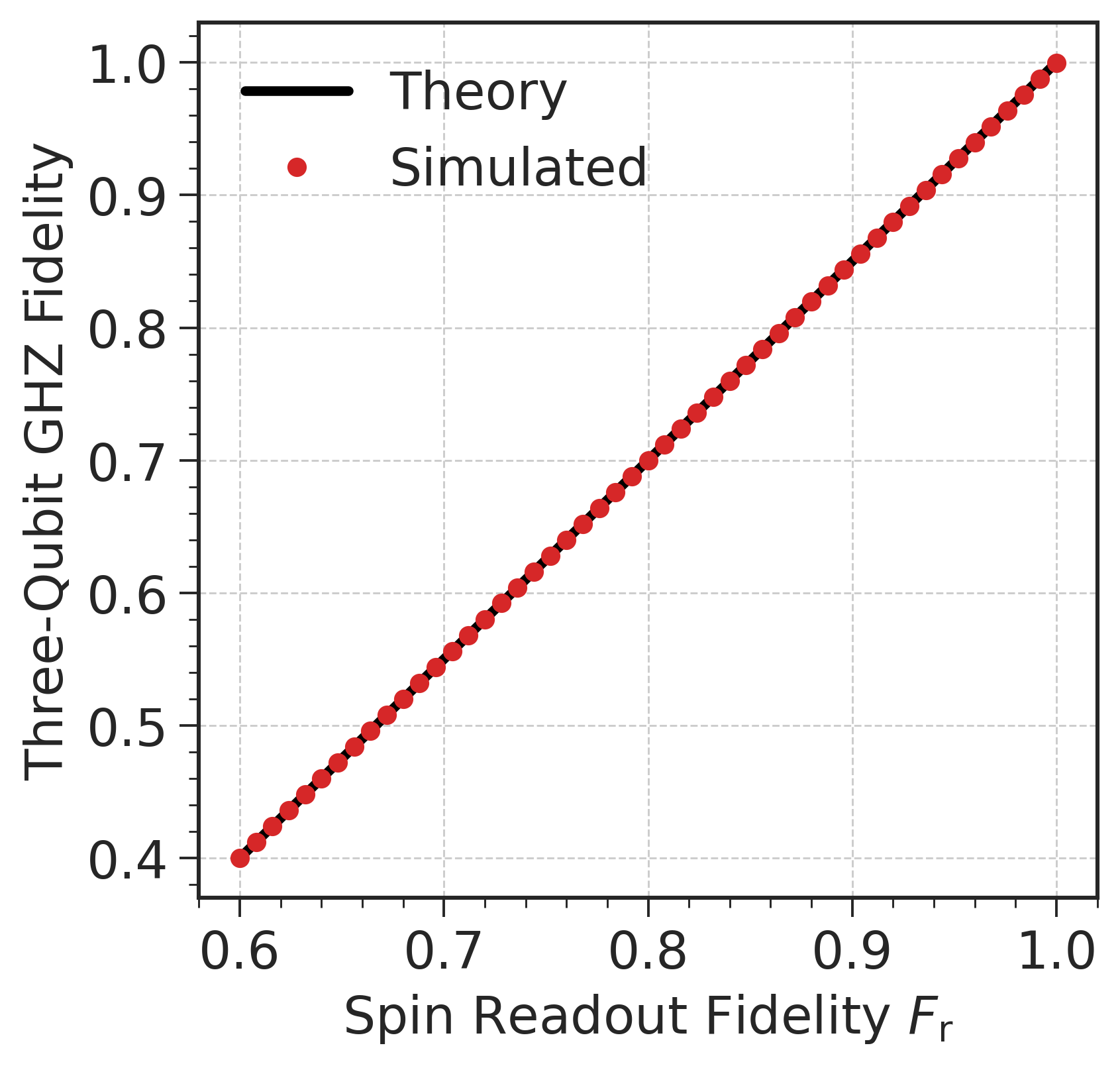}
    \caption{Comparison between the simulated and analytical fidelities under spin readout error.}
    \label{fig:compare_Readout}
\end{figure*}
Supplementary Fig.~\ref{fig:compare_Readout} shows great agreement between simulated fidelity and the analytical form in Eq.~(\ref{eq:Fread}).

\newpage
\subsection{Phonon-induced Pure Dephasing}

Unwanted coupling of the QD optical transition to a phonon thermal bath results in broadening of the QD zero-phonon line as well as a broad phonon sideband~\cite{besombes_2001_acoustic,krummheuer_2002_theory,muljarov_2004_dephasing,tighineanu_phonon_2018}. The former gives rise to dephasing of the excited state at a rate of $\gamma_d$ and subsequenct emission of incoherent photons,  which degrade their interference, whereas the latter is filtered out with frequency filters.
The effect of phonon-induced pure dephasing on the GHZ fidelity can be expressed by~\cite{tiurev_fidelity_2021}
\begin{align}
    \mathcal{F}^{(N)}_{\text{GHZ}} = \frac{1}{2}+\frac{1}{2}\bigg(\frac{\Gamma}{\Gamma+2\gamma_d}\bigg)^N,\label{eq:PureDephasing}
\end{align}
where $\Gamma$ is the total decay rate of the excited state $\ket{\Uparrow\downarrow\uparrow}$. To consider this in the Monte Carlo simulation, we introduce a random phase (0 to $2\pi$) between $\ket{\uparrow}$ and $\ket{\downarrow}$ after each photon generation event, with a probability determined by $\gamma_d/\Gamma$.
\begin{figure*}[h]
    \centering
    \includegraphics[scale=0.5]{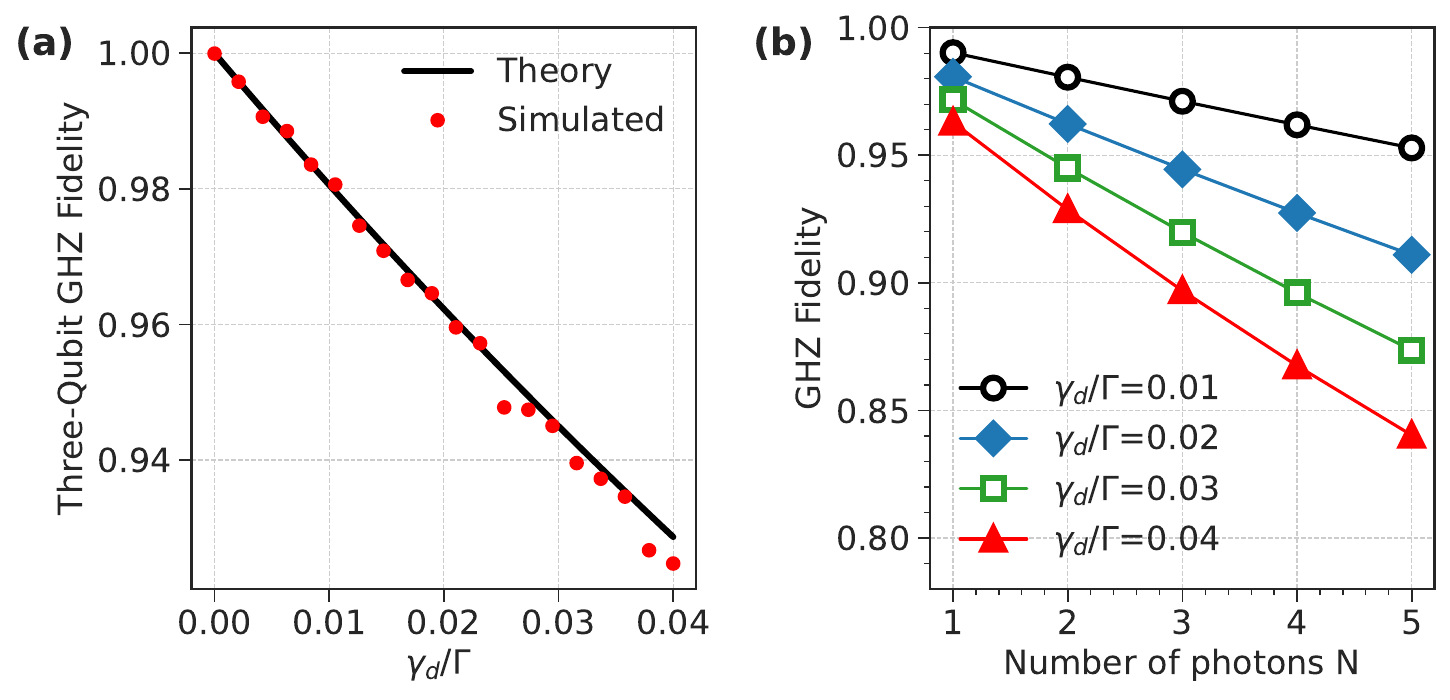}
    \caption{(a) Comparison between simulated and analytical fidelities under phonon-induced pure dephasing. (b) Plot of $\mathcal{F}^{(N)}_{\text{GHZ}}$ as a function of photon number $N$ for different values of the ratio between $\gamma_d$ and $\Gamma$.}
    \label{fig:compare_pure_dephasing}
\end{figure*}
\newpage
\subsection{Optical Excitation Errors}
In this section, we consider two error mechanisms that occur during optical pulsed excitation of the transition $\ket{\downarrow}\leftrightarrow\ket{\Uparrow\downarrow\uparrow}$. The first error originates from the spectral overlap between the excitation laser and the undesired transition $\ket{\uparrow}\leftrightarrow\ket{\Downarrow\downarrow\uparrow}$, which leads to a non-zero probability of exciting $\ket{\uparrow}$, emitting photons of a different frequency. Although these photons can be filtered out, they lead to dephasing of the coherence~\cite{tiurev_fidelity_2021}. 
The second error process takes place when the pulse duration of the excitation laser $T_{\text{pulse}}$ approaches the excited state lifetime $1/\Gamma$. In such a case, the QD could be driven twice emitting multiple photons within the excitation pulse.

To investigate the entanglement fidelity under these errors, for a single optical excitation in the Monte Carlo simulation, we first estimate the probabilities of off-resonant excitation (i.e., probability of driving $\ket{\uparrow}\to\ket{\Downarrow\downarrow\uparrow}$) and two-photon emission, which are found by solving optical Bloch equations describing the excitation and emission from a two-level system:
\begin{align}
 &
  \begin{cases}
    \dot{\rho}_{gg0} = \Omega(t)\cdot \mathrm{Im}(\rho_{eg0}) \\ 
     \dot{\rho}_{ee0}= - \Omega(t)\cdot \mathrm{Im}(\rho_{eg0})  - \rho_{ee0}\Gamma \\
     \dot{\rho}_{eg0} = i \cdot \Delta_L \cdot \rho_{eg0} + i \cdot \frac{ \Omega(t)}{2} \cdot (\rho_{ee0} - \rho_{gg0})- \rho_{eg0}\frac{\Gamma}{2}\\
  \end{cases}
  \label{eq:Excitation_error1}\\
  &
  \begin{cases}
    \dot{\rho}_{gg1} =  \Omega(t) \cdot \mathrm{Im}(\rho_{eg1}) + \rho_{ee0}\Gamma \\
     \dot{\rho}_{ee1} = - \Omega(t) \cdot \mathrm{Im}(\rho_{eg1}) - \rho_{ee1}\Gamma \\
     \dot{\rho}_{eg1} = i \cdot \Delta_L \cdot \rho_{eg1} + i \cdot \frac{ \Omega(t)}{2} \cdot (\rho_{ee1} - \rho_{gg1}) - \rho_{eg1}\frac{\Gamma}{2} \\
  \end{cases}
  \label{eq:Excitation_error2}\\
    &~~~\dot{\rho}_{gg2} = \rho_{ee1}\Gamma,\label{eq:Excitation_error3}
\end{align}
where $\rho_{ijk}$ is the atomic density matrix element $\ket{i}\bra{j}$, $i,j\in\{e,g\}$ and $k$ is the number of emitted photons considered up to 2 photons, i.e.,  $\rho_{gg0}$ ($\rho_{ee0}$) is the ground-state (excited-state) population in which no photon is emitted.  $\Omega (t)$ is the time-dependent optical Rabi frequency, assuming a Gaussian profile for the excitation laser. $\Delta_L$ is the laser detuning relative to an optical transition. $\Gamma$ is the excited-state decay rate assumed to be equal for $\ket{\Uparrow\downarrow\uparrow}$ and $\ket{\Downarrow\downarrow\uparrow}$.

To model the current experiment, we set the laser detuning $\Delta_L=$~\SI{12}{GHz} from the undesired transition $\ket{\uparrow}\to\ket{\Downarrow\downarrow\uparrow}$ to estimate the probability of off-resonant excitation, and $\Delta_L=$~\SI{2}{GHz} from the desired transition $\ket{\downarrow}\leftrightarrow\ket{\Uparrow\downarrow\uparrow}$ for computing the re-excitation probability. For the optimized InAs and GaAs QDs simulations, the corresponding $\Delta_L$ can be found in Supplementary table.~\ref{tab:simFidelityRealParmeter}. 
For simulations with a Gaussian excitation pulse, Eqs.~(\ref{eq:Excitation_error1})-(\ref{eq:Excitation_error3}) can only be solved numerically. Therefore, to verify our simulation code we now choose a square pulse and consider only the off-resonant excitation error, which enables a direct comparison with the closed form derived analytically in Ref.~\cite{tiurev_fidelity_2021}:
\begin{align}
    \mathcal{F}^{(N)}_{\text{GHZ}} = \frac{1}{2} \frac{D_1^{N} + D_2^{N}}{(D_2 + D_3)^{N}},
\label{eq:excitation_error}
\end{align}
where the relevant terms read
\begin{align*}
    D_1 &=   |c_0 c_2|^2 \\
    D_2 &=   |c_0 c_2|^2 + |c_0 \phi_2|^2  +  |\phi_0 c_2|^2+ |\phi_0 \phi_2|^2  \\
    D_3 &=   |c_1 c_3|^2 + |c_3 \phi_1|^2  + |c_1 \phi_3|^2 + |\phi_3 \phi_1|^2 + |c_3 c_2|^2 + |c_3 \phi_2|^2 + |\phi_3 c_2|^2 +|\phi_3 \phi_2|^2,
    \end{align*}
with the wavefunction coefficients
\begin{align*}
|\phi_1|^2 &= \frac{3 \sqrt{3}\pi}{8\tilde{\Delta}} - \frac{3\pi^2}{2\tilde{\Delta}^2} \left(\frac{3}{8} - \frac{1}{\pi^2}\right) 
, \, |\phi_2|^2 = \frac{\pi\sqrt{3}}{8\tilde{\Delta}} |c_2|^2
 , \,  |\phi_3|^2 = \frac{3}{16} \left(\frac{\sqrt{3}\pi}{8\tilde{\Delta}} - \frac{3\pi^2}{16\tilde{\Delta}^2}\right), 
|\phi_0|^2 = \frac{13 \sqrt{3}\pi}{128\tilde{\Delta}}|c_2|^2 \\
|c_0|^2 &= 1 ,\,  |c_1|^2 = \frac{\sqrt{3}\pi}{2\tilde{\Delta}}, \, |c_2|^2 = 1 - |c_1|^2, \, \tilde{\Delta}=\frac{\Delta_L}{\Gamma}.
\end{align*}
The expressions above assume perfect frequency filters and a square pulse of the optimal duration $T^{\text{opt}}_{\text{pulse}}=\sqrt{3}\pi/\Delta_L$ such that a optical $\pi$-pulse resonantly drives the transition $\ket{\downarrow}\leftrightarrow\ket{\Uparrow\downarrow\uparrow}$, while a $2\pi$-pulse is applied on the off-resonant transition.
We reproduce this condition in our simulation and compare it to the analytical form in Eq.~(\ref{eq:excitation_error}), see Supplementary Fig.~\ref{fig:compare_excitation_error}(a). We observe very good agreement between theory and simulation over a laser detuning from $\Delta_L=$~\SI{10}{GHz} to \SI{100}{GHz}. 
Supplementary Fig.~\ref{fig:compare_excitation_error}(b) shows the predicted scaling of the off-resonant excitation error with photon numbers.

\begin{figure*}[h]
    \centering
    \includegraphics[scale=0.5]{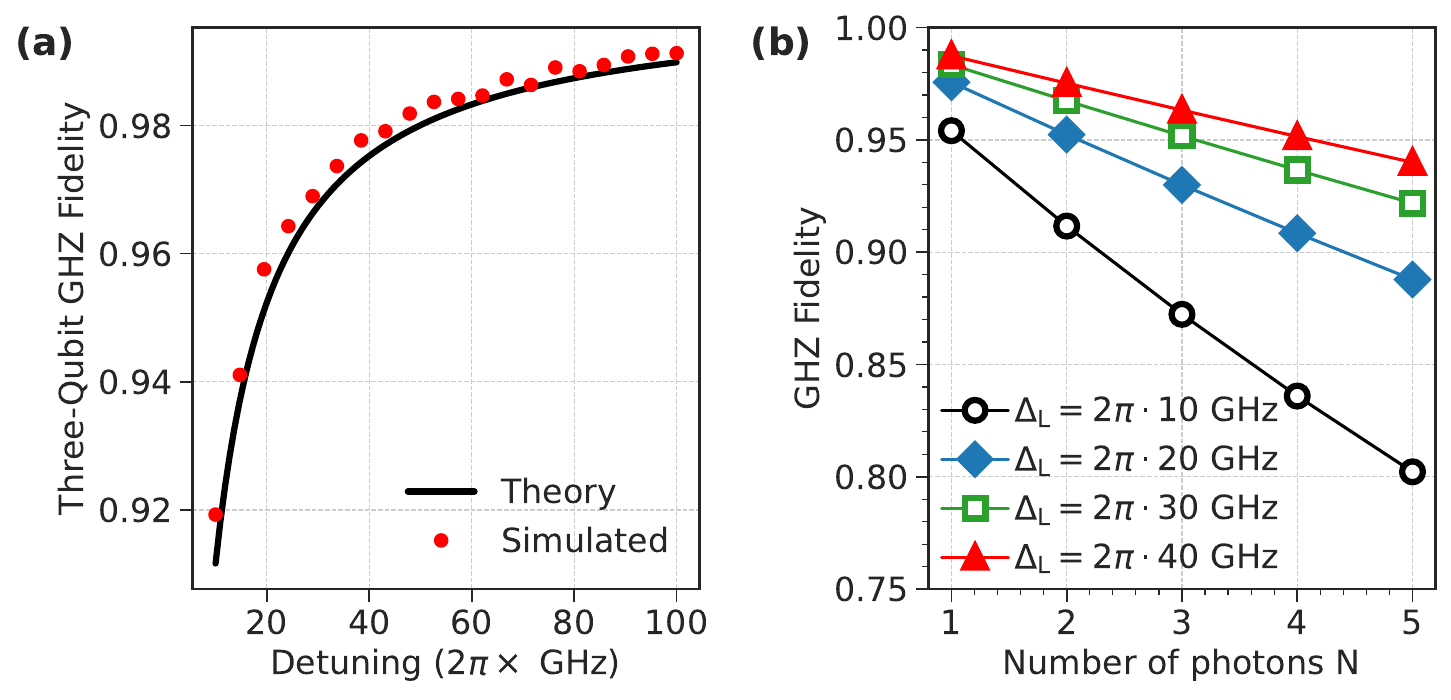}
    \caption{(a) Comparison between the simulated and analytical fidelities under off-resonant excitation error. (b) Plot of $\mathcal{F}^{(N)}_{\text{GHZ}}$ as a function of photon number $N$ for different values of the laser detuning $\Delta_L$. }
    \label{fig:compare_excitation_error}
\end{figure*}

\newpage
\section{Photon Detection Efficiency}

\begin{table}[htbp]

\begin{ruledtabular}
\renewcommand{\arraystretch}{1.3}

\begin{tabular}{ccccc}
\textbf{Loss source} & \textbf{Efficiency $\%$} & \textbf{Loss (dB)} & \textbf{Source}  \\
\hline
Waveguide-coupling factor $\beta$ &90&-0.46&Estimated\\
Less-than-$\pi$ excitation pulse &80&-0.97&Estimated\\

Emission into the zero phonon line&95&-0.22& Ref.~\cite{uppu2020_scalable} \\
Two-sided waveguide configuration &50&-3.01& Estimated\\
\begin{tabular}{@{}c@{}}Propagation loss in GaAs waveguide (free carrier absorption, grating coupler loss)\end{tabular}
  &56&-2.52& Measured \\
Transmission in the collection fiber&60&-2.22&Measured\\
\hline
\textbf{
\begin{tabular}{@{}c@{}}Total efficiency from quantum dot to the collection fiber\end{tabular} }&11.5&-9.4\\
\hline

\\
 50/50 Beamsplitter in TBI & 50&-3.01&Estimated\\

 Additional loss in TBI due to other optic elements & 60&-2.22&Measured\\

Total transmission after two etalon filters &90& -0.46& Measured\\
Fiber coupling to the detection fiber before SNSPDs & 70&-1.55&Measured\\
Fiber transmission to SNSPDs&85&-0.71&Measured\\
SNSPDs detection efficiency&70&-1.55&Measured\\
\hline
\textbf{State characterization and detection}&11.2&-9.5&\\
\hline
\textbf{Total loss from quantum dot to detection}&1.3&-18.9&\\
\end{tabular}
\end{ruledtabular}
\caption{\label{tab:table4} Compilation of all photon loss sources in the entanglement experiment. SNSPDs stands for superconducting nanowire single-photon detectors from PhotonSpot.}
\end{table}

\newpage

%


\bibliographystyle{apsrev4-1} 
\bibliography{Paper_3qubits_appendix} 